\documentclass[twoside,twocolumn,9pt]{article}
\usepackage{extsizes}
\usepackage[super,sort&compress,comma]{natbib} 
\usepackage[version=3]{mhchem}
\usepackage[left=1.5cm, right=1.5cm, top=1.785cm, bottom=2.0cm]{geometry}
\usepackage{balance}
\usepackage{times,mathptmx}
\usepackage{sectsty}
\usepackage{graphicx} 
\usepackage{amsmath,amssymb,bbm}
\usepackage{lastpage}
\usepackage[format=plain,justification=justified,singlelinecheck=false,font={stretch=1.125,small,sf},labelfont=bf,labelsep=space]{caption}
\usepackage{float}
\usepackage{fancyhdr} 
\usepackage{fnpos}
\usepackage[english]{babel}
\addto{\captionsenglish}{%

}
\usepackage{array} 
\usepackage{droidsans}
\usepackage{charter}
\usepackage[T1]{fontenc}
\usepackage[usenames,dvipsnames]{xcolor}
\usepackage{setspace}
\usepackage[compact]{titlesec}
\usepackage{hyperref}

\usepackage{epstopdf}

\definecolor{cream}{RGB}{222,217,201}

\usepackage{xspace}
\newcommand\mum{$\mu$m\xspace}
\usepackage{enumitem}

\usepackage{mathrsfs}

\begin{document}
  
\pagestyle{fancy}
\thispagestyle{plain}
\fancypagestyle{plain}{
\renewcommand{\headrulewidth}{0pt}
}

\makeFNbottom
\makeatletter
\renewcommand\LARGE{\@setfontsize\LARGE{15pt}{17}}
\renewcommand\Large{\@setfontsize\Large{12pt}{14}}
\renewcommand\large{\@setfontsize\large{10pt}{12}}
\renewcommand\footnotesize{\@setfontsize\footnotesize{7pt}{10}}
\makeatother

\renewcommand{\thefootnote}{\fnsymbol{footnote}}
\renewcommand\footnoterule{\vspace*{1pt}%
\color{cream}\hrule width 3.5in height 0.4pt \color{black}\vspace*{5pt}} 
\setcounter{secnumdepth}{5}

\makeatletter 
\renewcommand\@biblabel[1]{#1}            
\renewcommand\@makefntext[1]%
{\noindent\makebox[0pt][r]{\@thefnmark\,}#1}
\makeatother 
\renewcommand{\figurename}{\small{Fig.}~}
\sectionfont{\sffamily\Large}
\subsectionfont{\normalsize}
\subsubsectionfont{\bf}
\setstretch{1.125} 
\setlength{\skip\footins}{0.8cm}
\setlength{\footnotesep}{0.25cm}
\setlength{\jot}{10pt}
\titlespacing*{\section}{0pt}{4pt}{4pt}
\titlespacing*{\subsection}{0pt}{15pt}{1pt}

\fancyfoot{}
\fancyfoot[LO,RE]{\vspace{-7.1pt}\includegraphics[height=9pt]{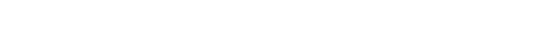}}
\fancyfoot[CO]{\vspace{-7.1pt}\hspace{13.2cm}\includegraphics{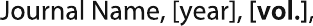}}
\fancyfoot[CE]{\vspace{-7.2pt}\hspace{-14.2cm}\includegraphics{RF}}
\fancyfoot[RO]{\footnotesize{\sffamily{1--\pageref{LastPage} ~\textbar  \hspace{2pt}\thepage}}}
\fancyfoot[LE]{\footnotesize{\sffamily{\thepage~\textbar\hspace{3.45cm} 1--\pageref{LastPage}}}}
\fancyhead{}
\renewcommand{\headrulewidth}{0pt} 
\renewcommand{\footrulewidth}{0pt}
\setlength{\arrayrulewidth}{1pt}
\setlength{\columnsep}{6.5mm}
\setlength\bibsep{1pt}

\makeatletter 
\newlength{\figrulesep} 
\setlength{\figrulesep}{0.5\textfloatsep} 

\newcommand{\topfigrule}{\vspace*{-1pt}%
\noindent{\color{cream}\rule[-\figrulesep]{\columnwidth}{1.5pt}} }

\newcommand{\botfigrule}{\vspace*{-2pt}%
\noindent{\color{cream}\rule[\figrulesep]{\columnwidth}{1.5pt}} }

\newcommand{\dblfigrule}{\vspace*{-1pt}%
\noindent{\color{cream}\rule[-\figrulesep]{\textwidth}{1.5pt}} }

\makeatother

\twocolumn[ 
  \begin{@twocolumnfalse}
  {\includegraphics[height=30pt]{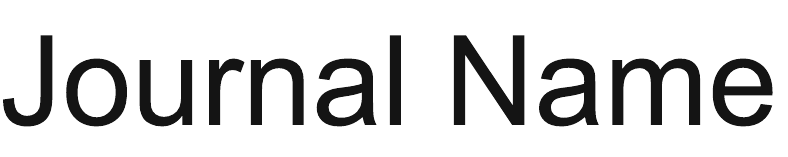}\hfill%
 \raisebox{0pt}[0pt][0pt]{\includegraphics[height=55pt]{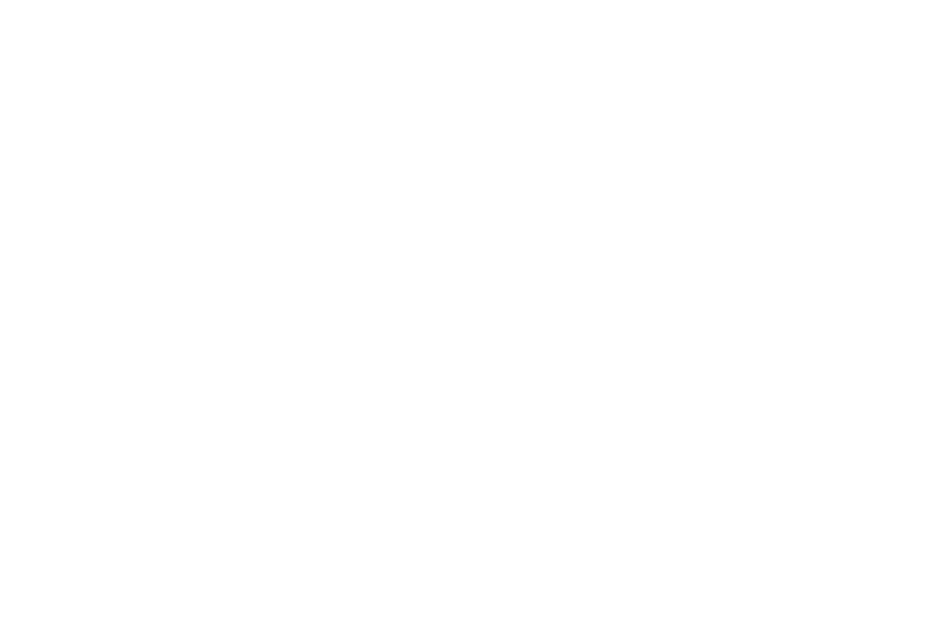}}%
 \\[1ex]%
}\par
\vspace{1em}
\sffamily
\begin{tabular}{m{4.5cm} p{13.5cm} } 

\includegraphics{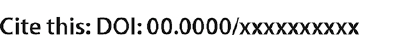} & \noindent\LARGE{\textbf{Swimming droplet in 1D geometries, an active Bretherton problem$^\dag$}} \\
\vspace{0.3cm} & \vspace{0.3cm} \\

 & \noindent\large{Charlotte de Blois,$^{a, b, \ddag}$ Vincent Bertin,\textit{$^{a, c, \ddag}$} Saori Suda,\textit{$^{d, \ddag}$} Masatoshi Ichikawa,\textit{$^{d}$} Mathilde Reyssat,\textit{$^{a}$} and Olivier Dauchot\textit{$^{a}$}} \\

\includegraphics{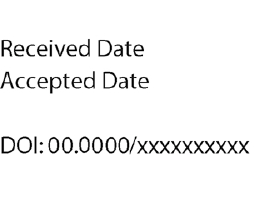} & \noindent\normalsize{We investigate experimentally the behavior of self-propelled water-in-oil droplets, confined in capillaries of different square and circular cross-sections. The droplet's activity comes from the formation of swollen micelles at its interface. In straight capillaries the velocity of the droplet decreases with increasing confinement. However at very high confinement, the velocity converges toward a non-zero value, so that even very long droplets swim. Stretched circular capillaries are then used to explore even higher confinement. The lubrication layer around the droplet then takes a non-uniform thickness which constitutes a significant difference with usual flow-driven passive droplets. A neck forms at the rear of the droplet, deepens with increasing confinement, and eventually undergoes successive spontaneous splitting events for large enough confinement. Such observations stress the critical role of the activity of the droplet interface on the droplet's behavior under confinement. We then propose an analytical formulation by integrating the interface activity and the swollen micelles transport problem into the classical Bretherton approach. The model accounts for the convergence of the droplet's velocity to a finite value for large confinement, and for the non-classical shape of the lubrication layer. We further discuss on the saturation of the micelles concentration along the interface, which would explain the divergence of the lubrication layer thickness for long enough droplets, eventually leading to the spontaneous droplet division. } \\%
\end{tabular}
\end{@twocolumnfalse} \vspace{0.6cm}
]

\renewcommand*\rmdefault{bch}\normalfont\upshape
\rmfamily
\section*{}
\vspace{-1cm}

\footnotetext{\textit{$^{a}$ UMR CNRS Gulliver 7083, ESPCI Paris, PSL Research University, 75005 Paris, France.}}
\footnotetext{\textit{$^{b}$ Okinawa Institute of Science and Technology Graduate University, Onna-son, Okinawa 904-0495, Japan}}
\footnotetext{\textit{$^{c}$ Univ. Bordeaux, CNRS, LOMA, UMR 5798, 33405 Talence, France}}
\footnotetext{\textit{$^{d}$ Department of Physics, Graduate School of Science, Kyoto University, Kitashirakawa-Oiwakecho, Sakyo-ku, Kyoto 606-8502, Japan.}}
\footnotetext{\dag~Electronic Supplementary Information (ESI) available: [(1) Drp\_in\_Cyl.avi (1~i/s~accelerated 20 times): video showing three experiments of droplets of different sizes swimming in circular capillaries of radius $h=50$ \mum. (2) Drp\_in\_Cons.avi (10~i/s~accelerated 30 times): video showing two experiments inside stretched capillaries of droplets simply elongating or spontaneously dividing while swimming.]. See DOI: 10.1039/cXsm00000x/}
\footnotetext{\ddag These authors contributed equally to this work} 


\section{Introduction}

Biological micro-swimmers exhibit a number of fascinating swimming strategies, to compensate for the absence of inertia. Even more intriguing is the way such organisms manage to probe and explore their environment, probing the presence of external fields such as temperature, nutriment concentration, gravity, etc. In many cases, they also manage to explore narrow channel-like passages, such as in soil~\cite{Or2007} or in the organism~\cite{Denissenko2012} of their host; or because they are placed in artificial micro-fluidic channels~\cite{Guasto2010, Pepper2010} to steer their motion~\cite{Liu1997,Das2015} with application in drug delivery. The euglenids~\cite{Noselli2019} is a striking example of such microorganisms, which are able to adapt its swimming strategy from flagellar propulsion to crawling. During this transition, the euglenids don't touch the wall, and are sensitive to the confinement through hydrodynamic interactions only. Another amazing example is that of paramecium~\cite{Mannik2009,Jana2012}, when they take a cylindrical shape to swim in narrow capillaries.

In the context of artificial micro-swimmers, a now classical strategy is to exploit phoretic effects~\cite{Anderson1989,Golestanian2007} to ensure propulsion by locally inducing gradients that generate a flow field around the swimmer, which in turn ensures its propulsion. The gradient can be induced by engineering an asymmetry in the swimming body -- the so called Janus particles -- and thereby obtain auto-phoretic swimmers (diffusio-phoresis~\cite{Howse2007a}, thermo-phoresis~\cite{Jiang2010}, electro-phoresis~\cite{Paxton2004}).
More recently, it was shown that a spontaneous symmetry breaking of the flow field, non linearly coupled to the advection-diffusion of the scalar field, can also lead to self-sustained propulsion~\cite{Michelin2013,Morozov2019}.
Swimming droplets, generating a solute gradient around them, are the prototypical experimental realization of this mechanism~\cite{Thutupalli2011,Izri2014,Maass2016,izzet2020}.

The presence of walls or obstacle can alter the swimming motion in different ways. The most common and unavoidable one is the disturbance of the hydrodynamic flow field. The case of weakly confined swimmers interacting with the boundaries only via the far-field hydrodynamics flow has been intensively studied theoretically~\cite{Lauga2009,Felderhof2010, Zottl2012a, Zhu2013, Acemoglu2014, Liu2014, Wu2015, Kanso2019}. In this situation, the flow field around the swimmer is affected through the no-slip condition at the boundaries. Theoretical studies for spherical swimmers~\cite{Zottl2012a,Zhu2013} demonstrated that the behavior of the swimmer (helical vs straight trajectory - attraction vs repulsion by the boundary) then strongly depends on its nature (pusher - puller - neutral). Experimental investigations on biological swimmers~\cite{Jeanneret2019} have revealed the diversity of the flow field developing around a micro-swimmer under such confinement. In the case of the phoretic swimmers, the transport of the scalar field will also be altered and thereby modify the swimming motion. The way a single flat boundary (a wall) alters the swimmer motion has been documented both theoretically and experimentally~\cite{Popescu2009, Chiang2014, Uspal2015, Ibrahim2016, Mozaffari2016}. In the case of the swimming droplets the solute is composed of micelles, the diffusion of which is way slower than molecular solutes. As a result, advection, which cannot be neglected in the transport of the solute, leads to a yet more complex situation because of the nonlinear coupling between the flow field dynamics and the advection-diffusion of the solute. The response of the swimmer motion to the proximity of the wall then depends on the relative importance of the advection and the diffusion of the scalar field. Quantitative measurements of the velocity field around a droplet swimming close to a wall could recently be analyzed and described theoretically~\cite{deblois2019}, but little is known about the swimming motion in more confined geometries such as micro-fluidic channels~\cite{Jin2017,Illien2020}.  

In the present work, we study experimentally the motion of a pure water swimming droplet~\cite{Izri2014}, in square and cylindrical capillaries with different levels of confinement. Amazingly, the droplet keeps its ability to swim under very strong confinement $L/2h = 10$, where $2h$ is the width of the capillary and $L$ is the length of the strongly elongated droplet. This is not only observed for square capillaries, but also for cylindrical ones, for which the droplet body is separated from the lateral boundaries by a lubrication film of a few microns in thickness. Furthermore for even larger confinement, we observe the spontaneous division of the droplet at its rear part. Both observations stress the crucial role of the active stresses at the droplet interface. The main goal of the present work is to quantify these new features, qualitatively different from that of flow-driven passive droplets and unveil the Marangoni-stress driven mechanisms coupled to the interface dynamics, responsible for them.

The paper is organized as follow. After a thorough description of the experimental setting, we characterize the swimming motion of the droplet in different channel-like geometries. We then propose a theoretical description, which accounts for the main observations, despite some important simplifications. Discussion about these simplifications and perspective for future investigations conclude the paper.

\section{Experimental setting\label{sec:ES}}

The experimental system is made of a water droplet inside a glass capillary,  filled with a continuous oil-surfactant phase, a squalane solution of mono-olein at a concentration $c=$ 25 mmol/L, that is far above the critical micellar concentration (CMC $\simeq$ 5 mmol/L).

In chambers~\cite{Izri2014} of diameter and thickness much larger than the droplet size and filled with the same oil-surfactant solution, such water droplets of typical size $a=100$ \mum spontaneously start swimming. The swimming motion results from the combination of two ingredients. First, the system is far from its physico-chemical thermodynamic equilibrium, which is a micro-emulsion made of inverse micelles filled with water, in the oil phase. As a result, a flux of water takes place continuously from the droplet to the inverse micelles~\cite{Herminghaus2014}. Secondly, the resulting isotropic concentration field of inverse swollen micelles happens to be unstable against an infinitesimal flow disturbance: in the presence of any tiny gradient of swollen micelles in the vicinity of the interface, Marangoni stresses and phoretic flows take place which induce a mobility of the droplet towards regions of small concentration, and therefore enhance the initial disturbance. For this instability to take place~\cite{Michelin2013,Izri2014,Morozov2019}, the P\'eclet number $Pe=U^* a/D$ must exceed some critical value $\text{Pe}_c=O(1)$, where $a$ is the radius of the droplet, $D$ is the diffusion coefficient of the micelles. and $U^*=\frac{A \mathscr{M}}{D}$ is the characteristic auto-phoretic velocity~\cite{Golestanian2007}, with $A$ the activity of the droplet and $\mathscr{M}$ the motility of the micelles. In other words for self-propulsion to occur, the diffusion of the micelles must be slow as compared to their advection by the Marangoni flow.

\begin{figure*}[t]
\includegraphics[width=0.7\textwidth]{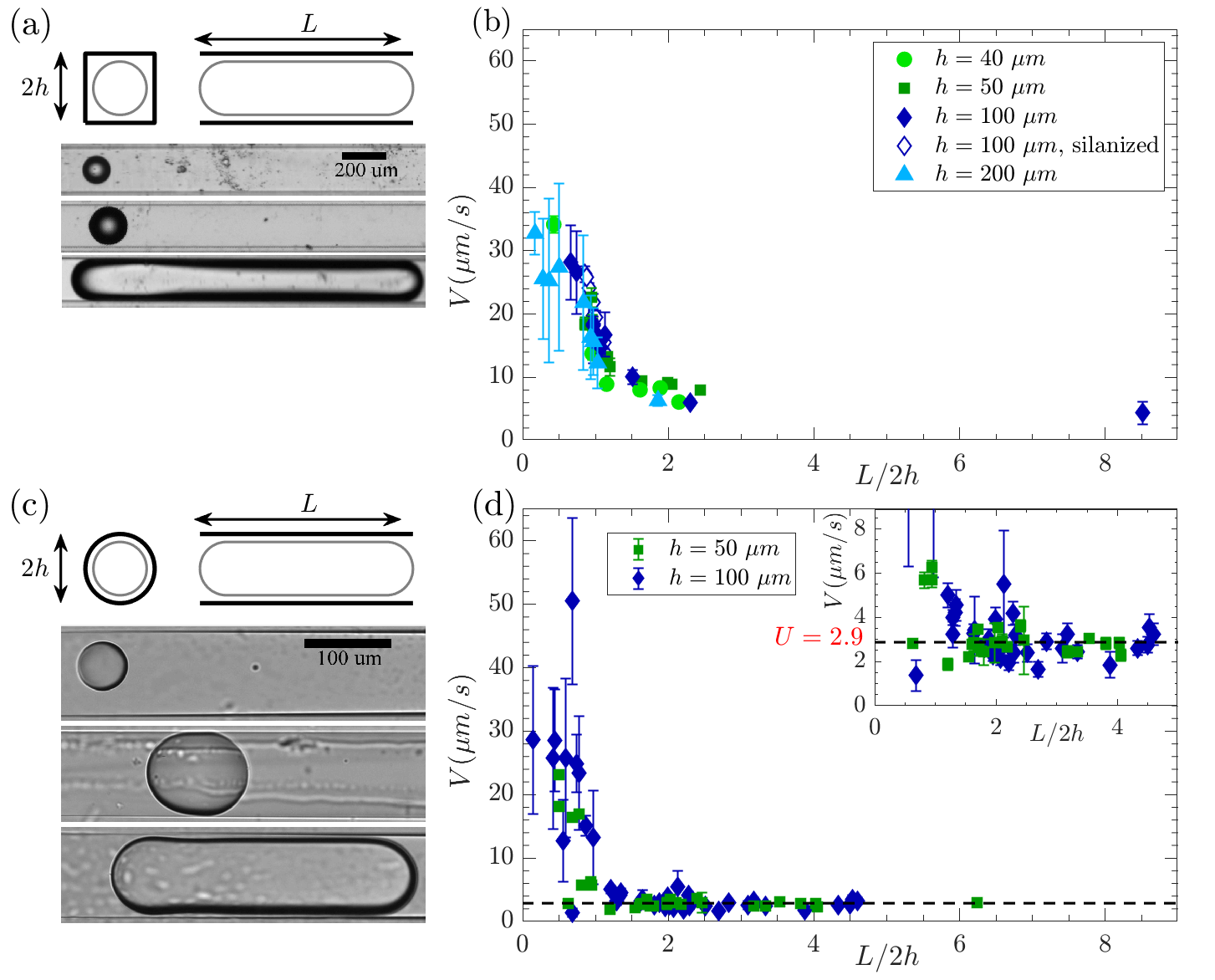}
\centering
1  \caption[Droplet velocity in square and cylindrical capillaries]{\textbf{{Droplet velocity in square (top) and cylindrical (bottom) capillaries}} (a) Sketch of a droplet in a square capillary, with snapshots of three droplets under increasing confinement $\frac{L}{2h}=0.5$, $\frac{L}{2h}=1$, $\frac{L}{2h}=8.5$ in a square capillary of half-width $h=100$ $\mu$m. (b) Velocity $V$ of droplets of various lengths $L$ swimming alone in square capillaries of different half-heights $h$, as a function of the confinement $\frac{L}{2h}$. (c) Sketch of a droplet in a circular capillary, and snapshots of three droplets under increasing confinement $\frac{L}{2h}=0.5$, $\frac{L}{2h}=1.5$, $\frac{L}{2h}=4$ in a circular capillary of radius $h=50$ $\mu$m. (d) Velocity $V$ of droplets of various lengths $L$ swimming alone in circular capillaries of different radii $h$, as a function of the confinement $\frac{L}{2h}$. The straight horizontal dotted line stresses the fact that even for the largest confinement, the droplet velocity is non-zero. The inset is a zoom of the same data on the low velocity values.}
\label{fig:V}
\end{figure*}

Here we confine such droplets in micro-channels, of typical length $\sim 2$cm much longer than the droplet size and with different cross-sectional geometries of typical inner size $h$ in the range 40\mum$<h<$200\mum, comparable to or smaller than the droplet size. Three different 1D geometries are used: square glass capillaries (Figure~\ref{fig:V} (a)), $h$ is then defined as half the inner dimension of the capillary, circular glass capillaries (Figure~\ref{fig:V} (c)), $h$ is then defined as the radius of the capillary, and stretched circular capillaries (Figure~\ref{fig:El}) whose inner radius varies continuously along their length between $h=100$ \mum (at both ends), and a constriction of radius $h_\text{min}$ that ranges from 30 \mum to 80 \mum, in the middle of the capillary, with a typical gradient of diameter $\frac{d h}{d x}=\pm 0.02$. Then relative to the swimming of the droplet, these capillaries present a convergent region followed by a divergent one.

At the beginning of each experiment, one droplet is produced at one end of a capillary previously filled with the oil-surfactant solution. The droplet spontaneously starts swimming. Both capillary ends are left open to the air. No external flow is imposed, and we ensure that there is no global flow by checking that the oil-air interface is not moving. This will be confirmed in section~\ref{sec:PIV}, through flow field measurements. During the experiment, the droplet swims from one end of the capillary to the other in typically one hour. Three sets of experiments are conducted. The first set focuses on the shape detection and the tracking of the droplet in square, circular and stretched capillaries (section~\ref{sec:V} and~\ref{sec:V2}), with an image acquisition rate of \mbox{$f_\text{acq}=1$ Hz}. A second set of experiments is dedicated to the study of the flow field around the droplet in circular capillaries using particle image velocimetry (PIV) (section~\ref{sec:PIV}). The image acquisition frequency is then $f_\text{acq}=10$ Hz. For the third set of experiments, a high-speed camera is used to capture the dynamics of the rear of the droplet in stretched capillaries during splitting events (section~\ref{sec:Dyn}), with an acquisition rates $f_\text{acq}=1$ kHz or $f_\text{acq}=10$ kHz. A complete description of the materials and methods is given in the appendix~\ref{sec:MM}.

\section{Experimental results}

We start by conducting experiments in glass capillaries of square or circular cross-section of comparable inner size between $h=40$~\mum and $h=200$~\mum.  
 Upon production, in square capillaries, all the droplets start swimming. In circular ones, droplets of size up to six times the capillary diameter also swim. We observe that the droplets produced with longer sizes spontaneously divide into two droplets which each starts swimming in opposite directions: the one swimming toward the closest end of the capillary immediately gets stuck on the oil-air interface, while the other swims until the other end of the capillary. After less than a minute, all swimming droplets reach a stationary state, and keep a persistent direction with the exception of the very small droplets, essentially unconfined ($\frac{L}{2h}<0.2$), that sometime change direction. In the following, we focus on droplets swimming persistently in one direction.

\subsection{Droplet Velocity vs. Confinement\label{sec:V}}

\begin{figure*}[t]
\includegraphics[width=0.9\textwidth]{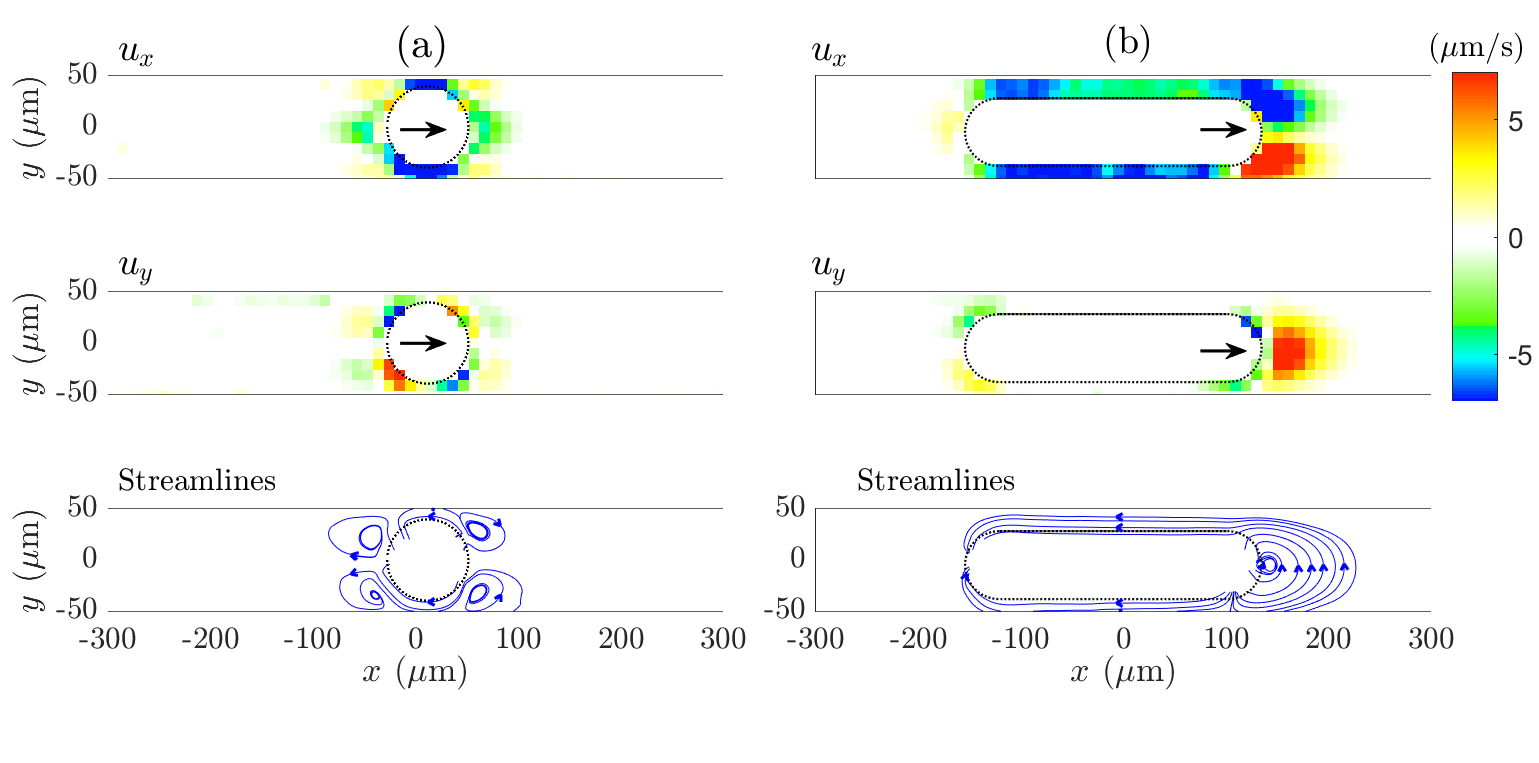}
\centering
\caption[PIV of the flow field around droplets swimming in circular capillaries]{\textbf{PIV of the flow field around droplets swimming in circular capillaries}: PIV around (a) a circular droplet ($\frac{L}{2h}=1$) and (b) a long droplet ($\frac{L}{2h}=3$) in a circular capillary of radius $h=50$ $\mu$m. The two first rows display the color-map of the velocities $u_x$ (direction of swimming) and $u_y$ around the droplet in the reference frame of the lab. The black arrows point the swimming direction of the droplet. The third row shows several streamlines around the droplet, with blue arrows that show the local direction of the flow. The inside of the droplet, masked during the PIV, is delimited by black dashed lines. $x$ is the direction of swimming of the droplet, $(xy)$ is the visualisation plane, orthogonal to the direction of the gravity $z$. The droplet's mask used for the PIV analysis is slightly smaller than the droplet size (for PIV requirement). As a result the gap between the drop and the wall is seen larger than in reality in this representation. The flow field is to be understood as being integrated over the thickness of the PIV depth of field of a few microns.}
\label{fig:PIV}
\end{figure*}

The behaviour of a swimming droplet strongly depends on the confinement, but is similar for different capillary geometries. A video of the swimming of three droplets of different sizes ($L=50$~\mum, 90 \mum or 400 \mum) in circular capillaries of radius $h=50$ \mum is provided in the electronic supplementary information. Typical shapes of droplets in square and circular capillaries are shown in Figure~\ref{fig:V}, together with the dependence of their velocity with the confinement $\frac{L}{2h}$. For all droplets, we measure the velocity averaged in time $\langle V \rangle$, together with its standard deviation $\sqrt{\langle V^2 \rangle - \langle V \rangle ^2}$.
 
Droplets smaller than the capillary inner size are spherical of diameter $L$ ($\frac{L}{2h}<1$). Such droplets oscillate between the two sides of the channel (in the median plane perpendicular to gravity). Because of buoyancy, they follow the bottom wall, thus performing a 2D motion on the bottom plane of a square capillary, and a 3D motion on the curved bottom surface of a circular capillary. The velocity of the droplet varies during an oscillation: it is minimum when the droplet reaches a side wall, and maximum far from the walls. The variations of the velocity during an oscillation period explain the large standard deviation of the velocity. Existing theoretical works~\cite{Zottl2012a,Zhu2013} on the behavior of swimmers in channels of circular section have predicted the existence of helical trajectories for neutral squirmers. Identifying precisely whether the observed oscillations pertains to this class of dynamics would require extracting the 3D trajectories of the droplets, together with their surrounding flow field, an interesting perspective for future studies.

Droplets larger than the capillary inner size take an elongated shape ended with two spherical caps. The total length of the droplet is denoted $L$ ($\frac{L}{2h}\geq1$). The elongated part of the droplet is separated from the glass wall by a film of oil (the droplet is never observed to wet the glass). Inside a square capillary, the gutters of the square cross-section give space for the outer fluid to flow. Inside a cylindrical capillary, the droplet is separated from the wall by a lubrication film of up to a few micro-meters in thickness only. We observe that the cross-section of the droplet has a radius that varies along its length, reaching a local minimum at the rear of the droplet. The region where the droplet radius is the thinnest is called the neck in what follows. 

As expected, the velocity decreases with increasing confinement. More surprisingly, for confinement greater than 1 (elongated droplets), the droplet velocity rapidly converges toward a small but \emph{finite} value. As a matter of fact, the velocity remains constant while further increasing the confinement, up to $\frac{L}{2h}=9$ in square capillaries and $\frac{L}{2h}=6.5$ in circular capillaries. This is all the more intriguing in the case of the cylindrical capillaries, for which the mass conservation imposes that a significant amount of fluid must be driven through the thin lubrication film, leading to a potentially strong dissipation. 

In the following, we focus on the behaviour of long droplets ($\frac{L}{2h}>2$) in circular capillaries, the most intriguing situation and also the simplest geometry to handle theoretically. 

\subsection{Flow field\label{sec:PIV}}

\begin{figure*}[t]
\includegraphics[width=0.9\textwidth]{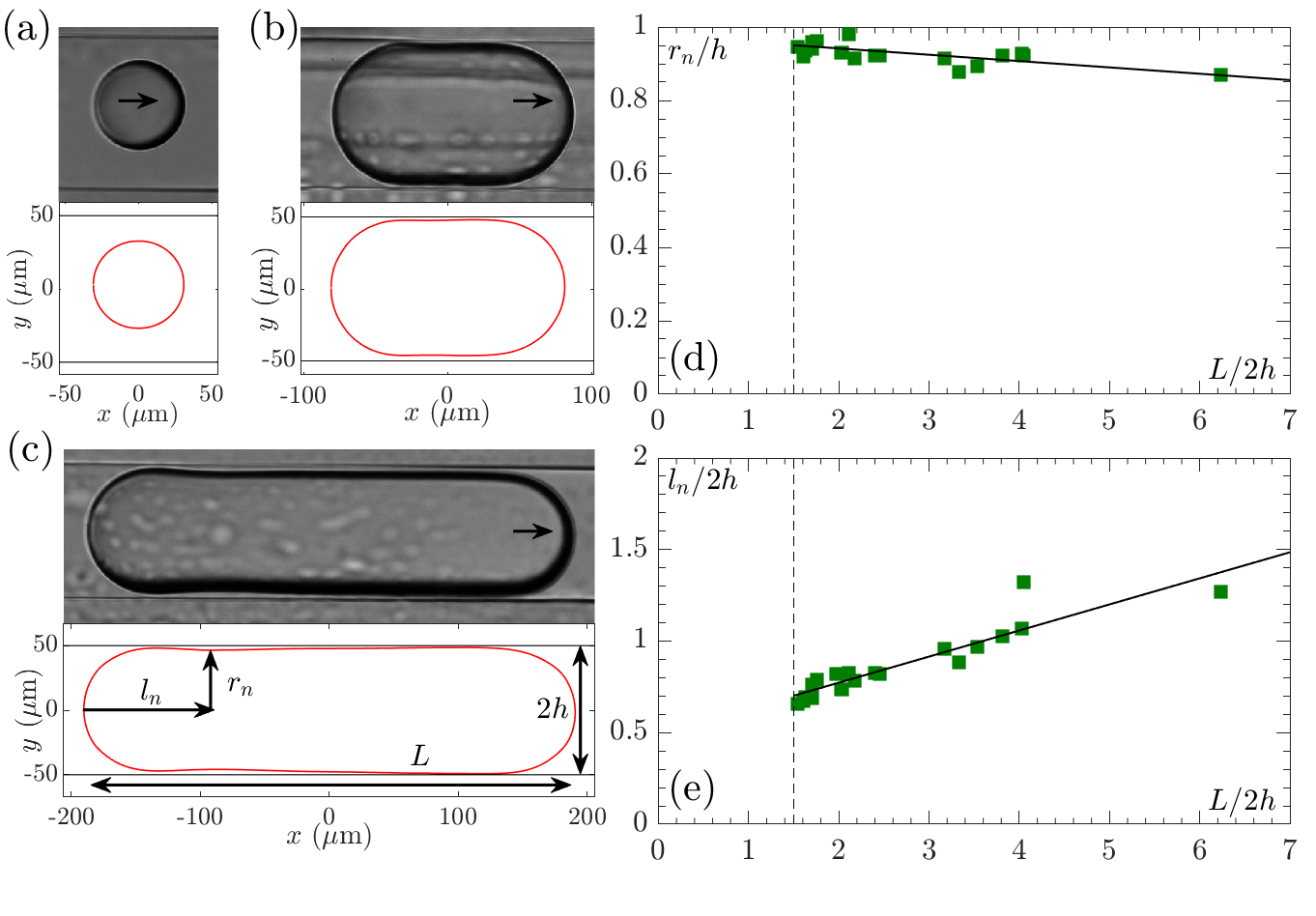}
\centering
\caption[Shape of an elongated droplet]{\textbf{Shape of an elongated droplet}: (a), (b) and (c), image and corresponding shape of droplets of size (a) $\frac{L}{2h}=0.5$, (b) $\frac{L}{2h}=1.7$, (c) $\frac{L}{2h}=4$. Long droplets present a neck of radius $r_n$ located at a position $l_n$ from the back of the droplet. (d) Evolution of the dimensionless neck radius $\frac{r_n}{h}$ with the confinement $\frac{L}{2h}$. (e) Evolution of the dimensionless neck position $\frac{l_n}{2h}$ with the confinement $\frac{L}{2h}$. The continuous black lines correspond to a linear regression of the data.}
\label{fig:Shape}
\end{figure*}

We start with PIV measurements of the flow field around the droplet. 
PIV is performed in the median plane, perpendicular to the gravity, of droplets swimming in cylindrical capillaries of radius $h=50$. Figure~\ref{fig:PIV} displays, from top to bottom, the component $u_x$ and $u_y$ of the flow field and a few selected streamlines around (a) a spherical droplet of typically the size of the capillary, $\frac{L}{2h}=1$, and (b) a long droplet of size $\frac{L}{2h}=3$. 
Note that, although the $u_x$ velocity component, strictly at the apex of the droplet, is expected to be positive and equal to the droplet velocity, we measure a slightly negative value for $u_x$ in front of the droplet. It is likely that we do not resolve well enough the velocity field at the interface of the droplet because of the conjugated effects of the non-zero thickness of the illumination plane, the resolution of the PIV, and the 3D recirculation flow that takes place close to the interface. This observation calls for a careful interpretation of our PIV observations, which should be taken as a qualitative image of the flow field, and not a quantitative 3D description as done by some of us in~\cite{deblois2019}, when characterizing the dynamics of droplet swimming above a bottom wall. Similarly the resolution of the PIV (10 \mum/pixel), is not large enough to resolve the flux in the lubrication film on the side of the droplet, and therefore only gives a global indication of the flow direction,thanks to the depth of field of the PIV setting.

This being said, the first crucial observation is that there is no flow far from the droplet (although the two ends of the capillary are left open to the air). All flows take place close to the droplet interface, up to a distance of typically the droplet size.  

In the case of spherical droplets, the flow is symmetric in $y$, and two recirculating regions are observed at the front and back of the droplet. The velocity of the fluid in these regions is similar in magnitude to the droplet velocity ($\sim 3$ \mum/s). It is also worth noting that the dominant symmetry of the flow field around the droplet is quadripolar, in contrast with the flow field aroung 3D unconfined squirmers, where the dipolar symmetry is dominant, and with that of swimming droplets close to a wall~\cite{deblois2019}, where the monopolar symmetry is dominant.

In the case of long droplets, a large recirculating region is observed at the front of the droplet, breaking the symmetry along the $y$ axis. The direction of the recirculation is stable during one experiment but switches between the two possible directions from one experiment to another.  The physical origin of the recirculation remains unclear, but it may result from an instability of the stagnation point of the flow at the apex of the droplet. The velocity of the fluid in this region ($u_x\sim 10$ \mum/s) is larger than the droplet velocity ($U\sim 3$ \mum/s). Finally, we notice that the PIV at the back of the droplet is disturbed by an agglomeration of the tracers at a stagnation point located at the extremity of the back cap, which makes the precise flow profile not fully resolved in this region. One can still observe a strong asymmetry between the front and rear in the amplitude of the flow, contrasting with the case of the spherical droplets. A more advanced interpretation of the above descriptions would require a better knowledge of the flow field around a confined active droplet, which is not the purpose of this paper, and is left for future work.

\subsection{Shape of the long droplets}

Typical shapes for different droplet sizes are illustrated in Figure~\ref{fig:Shape}, (a), (b) and (c). The shape of a droplet is stable and averaged over the duration of the experiment. Droplets smaller than the capillary diameter ($\frac{L}{2h}$) are spherical (a) while longer droplets take an elongated shape ((b) and (c)). This elongated shape present two particularities. First the thickness of the lubrication film is not constant, but increases toward the back of the droplet; this is a tiny but systematic effect. Second, for confinement larger than $\frac{L}{2h}=1.5$, a neck appears at the rear of the droplet. Figure~\ref{fig:Shape} (d) and (e) display the neck radius $r_n$, and its distance from the back of the droplet $l_n$. At the smallest confinement, the neck forms at a distance $\frac{l_n}{2h}\simeq0.5$. The neck is then shallow, $\frac{r_n}{h}\lesssim 1$. As the confinement increases, the neck goes further away from the rear of the droplet, and deepens. The shape of the droplet rear hence depends on the droplet length. 

The fact that the film thickens from the front to the rear of the droplet and that the neck shape depends on the droplet length contrast with the standard Bretherton phenomenology~\cite{Bretherton1961}, which describes the shape and motion of passive droplets driven externally. This underlines the conceptual difference with the present case, where active droplets are driven by self-induced local flows and calls for investigation at even higher confinement.

\begin{figure}[t]
\includegraphics[width=0.9\columnwidth]{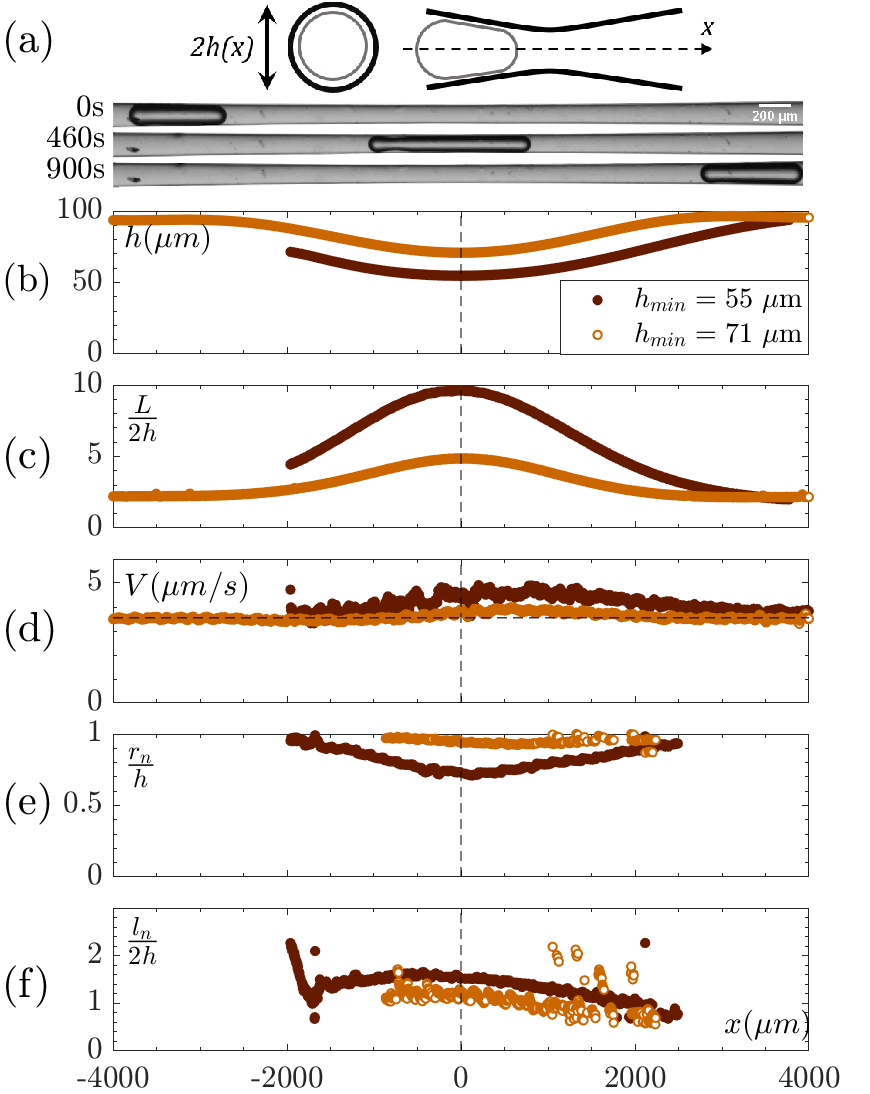}
\centering
\caption[Elongation of a swimming droplet in a stretched capillary]{\textbf{Elongation of a swimming droplet in a stretched capillary}: (a) Sketch of the geometry and images of the droplet at three different times corresponding to three positions in the stretched capillary. At 460s, the droplet is at the position where the confinement is the highest. (b) to (f) provide the evolution of several quantities with the position $x$ of the droplet in the capillary. (b) Height of the capillary at the center of mass of the droplet, (c) confinement of the droplet, (d) velocity of the droplet, (e) dimensionless radius of the neck, (f) dimensionless distance of the neck from the front of the droplet. The vertical black dashed line marks the position of the minimum capillary diameter. The two colors correspond to two different experiments made with different droplet sizes and capillary shapes.}
\label{fig:El}
\end{figure}
 
\begin{figure}[t]
\includegraphics[width=0.9\columnwidth]{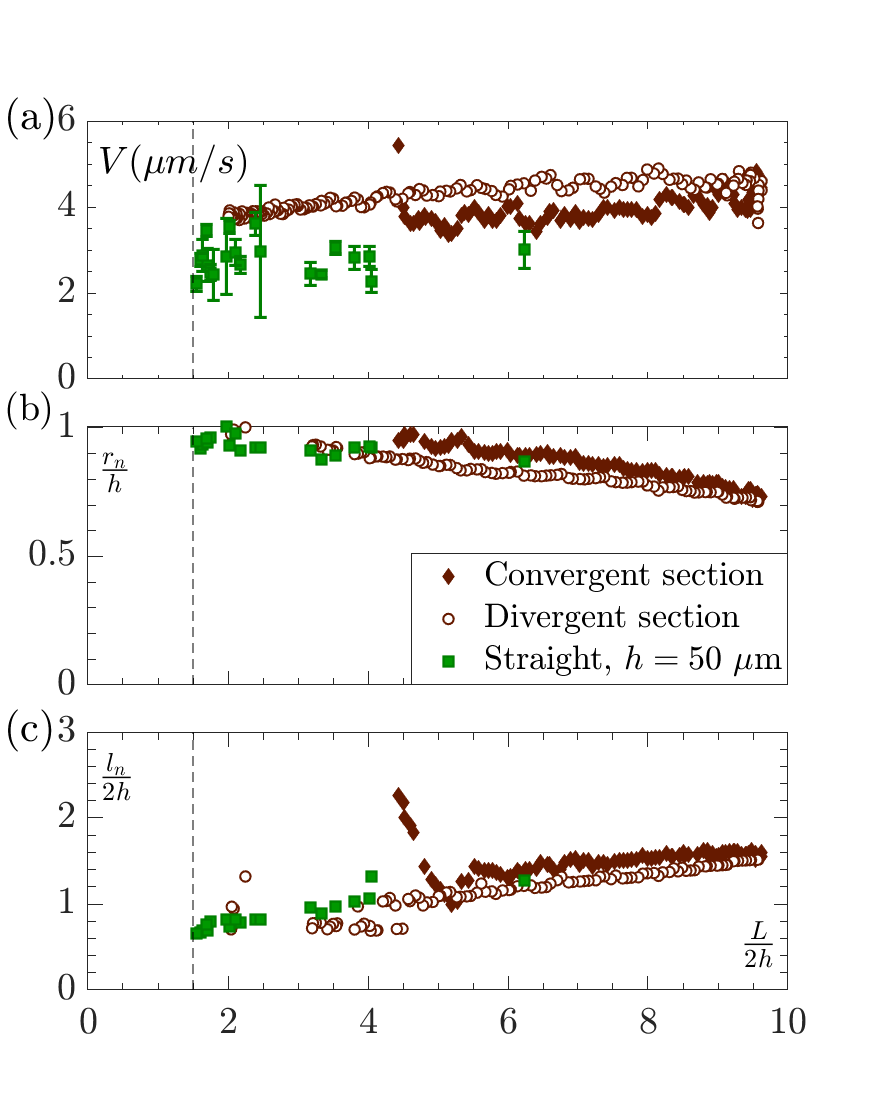}
\centering
\caption[Convergent vs. Divergent dynamics]{\textbf{Convergent vs. Divergent dynamics}: Evolution of (a) the droplet velocity, (b) the dimensionless neck radius and (c) the dimensionless neck position with the confinement; for droplets in straight circular capillaries of half height $h=50$ $\mu$m (green squares), for one droplet in a stretched capillary, in the convergent section (dark brown filled diamond), and in the divergent section (dark brown empty circles).}
\label{fig:Neck}
\end{figure}

\subsection{Further confinement\label{sec:V2}}
 
Very long droplets ($\frac{L}{2h}>7$) in circular capillaries are not stable at production and spontaneously divide into two or more droplets. A way to explore higher confinement and to probe continuously its effects on the droplet behavior, is to conduct experiments in stretched circular capillaries. Such devices keep the ideal circular cross-section and mimic perfectly situations where real swimmers have to experience confinement gradient.

\begin{figure}[t]
\includegraphics[width=0.9\columnwidth]{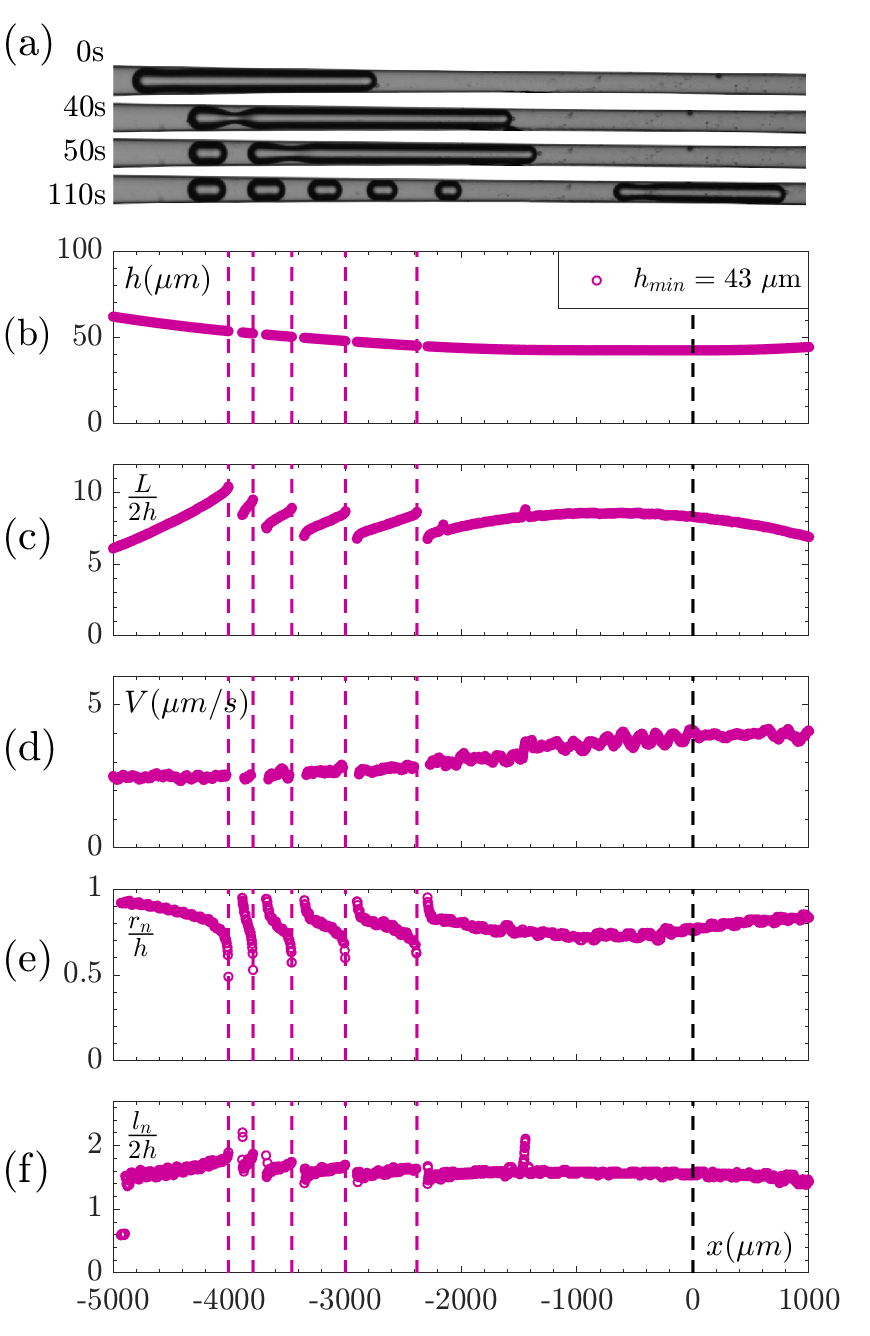}
\centering
\caption[Behaviour of a swimming droplet in a stretched capillary - division]{\textbf{Division of a droplet in a stretched capillary}: (a) images of a droplet at different positions in the stretched capillary: successive divisions occur. (b) to (f) give the evolution of quantities depending of the position $x$ of the droplet in the capillary, for the corresponding experiment. (b) Height of the capillary at the center of mass of the droplet, (c) droplet confinement, (d) droplet velocity, (e) dimensionless radius of the neck, (f) dimensionless distance of the neck from the front of the droplet. The pink dashed lines correspond to a division. The black dashed line corresponds to the position of the minimum height in the capillary.}
\label{fig:Div}
\end{figure}

Upon production at one end of the capillary ($h=$50~\mum), the droplet (of typical length between 50 and 150 \mum) starts swimming. As it swims down the convergent part of the capillary, the local radius of the capillary decreases and the length of the droplet increases per conservation of the droplet volume $\mathcal{V}$ ($L\propto\frac{\mathcal{V}}{h^2}$), as does the confinement ($\frac{L}{2h}\propto\frac{\mathcal{V}}{h^3}$). Two different features are observed, a simple elongation of the droplet, followed by a contraction after the constriction, or spontaneous division, presented respectively in Figures~\ref{fig:El} and~\ref{fig:Div}. A video of the swimming of two droplets in the two cases is provided in the electronic supplementary information. For each experiment, the droplet is tracked, and its shape detected along its motion. At each time, the radius of the capillary $h$ at the position of the droplet center of mass (b), the confinement of the droplet $\frac{L}{2h}$ (c), the velocity of the droplet $V$ (d), the dimensionless neck radius $r_n$ (e) and the dimensionless neck position $l_n$ (f) are measured as a function of the position of the droplet center of mass $x$. 

Let us first focus on the simple elongation of the droplet. Figure~\ref{fig:El} shows in (a) three snapshots of a droplet swimming in a stretched circular capillary, when it is in the convergent region, at the constriction, and when it is in the divergent region. Two different experiments are shown in~Figure~\ref{fig:El}, corresponding to different stretched capillaries, with a minimal radius of $h_\text{min}=55$\mum and $h_\text{min}=71$\mum respectively. Throughout the experiment, the droplet swims from one end of the capillary to the other. We note a small variation of the droplet velocity and a more significant dependence of the neck position and radius with the confinement. More importantly, although the thickness profiles of the stretched capillary in the converging and diverging regions are symmetric, the shape and speed of the droplet are not. This is further enlighten on Figure~\ref{fig:Neck}, where we compare the dependence on the confinement of (a) the velocity of the droplet, (b) the neck radius and (c) the neck position between the convergent and divergent part of the capillary for the constriction, $h_\text{min}=55$ \mum (brown diamonds and circles) and the straight circular capillaries (green squares).

The velocity slightly increases with the confinement, but also presents an hysteresis between the convergent (brown diamonds) and divergent (brown circles) regions of the capillary. This variation is most likely due to the gradient of capillary radius and would also exist for passive droplets : the difference of curvature between the front and back meniscus induces a capillary-induced pressure gradient inside the droplet, which, for passive droplets, makes the droplet move toward the highest radius. For an active droplet, this effect slows down the droplet in a convergent tube, while it accelerates it in a divergent one. This effect remains however weak, and comparable in magnitude to the variability of the velocity from one experiment to another (see inset of Fig~\ref{fig:V}.d).

Figure~\ref{fig:Neck} (b) and (c) exhibits the dependence of the shape of the droplet rear with the confinement : the radius of the neck decreases linearly, and its position goes further away from the back of the droplet, for increasing confinement. This evolution is reversible when the confinement is decreasing in the divergent region, indicating that the dynamics can safely be considered quasi-static, and that the influence of the capillary-induced pressure gradient is not significant here.
Finally, further increasing the confinement, the neck is expected to narrow, until the droplet eventually divides spontaneously.

\begin{figure}[t]
\includegraphics[width=0.85\columnwidth]{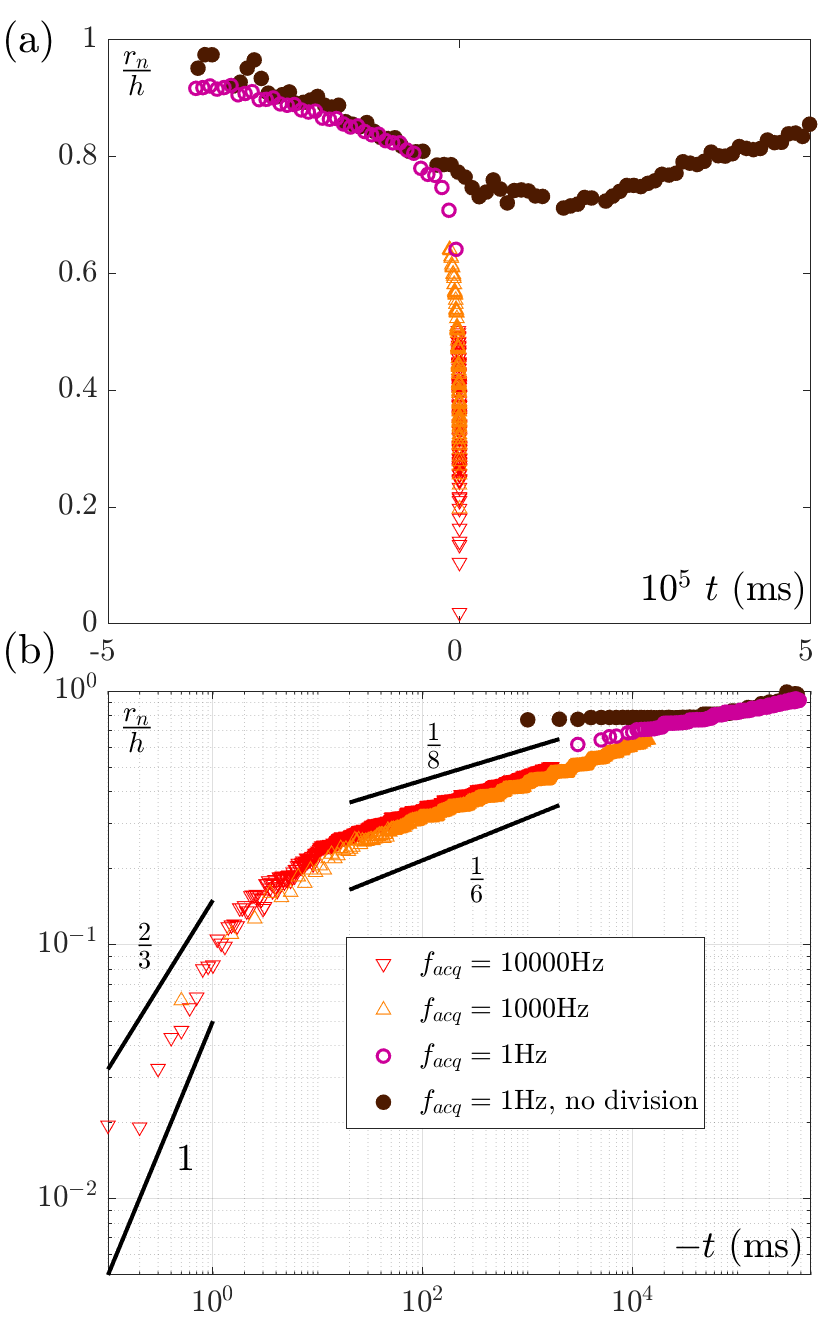}
\centering
\caption[Dynamic evolution of the neck]{\textbf{Dynamic evolution of the neck}: (a) linear plot and (b) logarithmic plot of the evolution of the dimensionless neck radius with time. The red and orange triangles and the pink empty circles correspond to experiments at different acquisition frequencies of droplet dividing, considering only the data up to the first division. $t=0$ corresponds to the time of division. The brown full circles correspond to an experiment with no division, $t=0$ is then chosen so that the convergent part of the experiment aligns with the others. The continuous black lines provide a lower and upper bound for the power law behaviors of the two fast regimes.}
\label{fig:Dyn}   
\end{figure}

\subsection{Spontaneous division of the droplet\label{sec:Dyn}}

The spontaneous division of droplets has been observed systematically for a dozen of different experiments with a number of successive divisions ranging from one to fourteen. For the sake of conciseness, the results in this section are presented in Figure~\ref{fig:Div} using the data from one typical experiment, but all following observations are valid for all experiments. At first the droplet swims and elongates as it moves toward the convergent region of the capillary. This behavior is similar to the one observed in the previous experiment described in Figure~\ref{fig:El}. When the droplet becomes "too confined" (at time $t=40$s in the presented experiment and in the supplementary video), it undergoes a spontaneous division at the position of the neck, as can be seen in Figure~\ref{fig:Div} (a). The daughter droplet (formed by the previous rear of the droplet) does not swim, which suggests that there is no fuel anymore for propulsion, namely that all micelles present in its environment have been saturated with water. 
The main droplet that shrunk in volume keeps swimming in the same direction. As the confinement of the droplet further increases, the droplet eventually divides a second time or more. Once the main droplet reaches the divergent region, its length decreases due to volume conservation. No division is observed in the divergent region. 
The behaviour of the droplet only differs from the simple elongation case at the approach of a division event ($\sim50$ s before division). 

Let us describe a succession of spontaneous divisions (Figure~\ref{fig:Div} (b) to (f)). Right before the division, the neck rapidly shrinks (e), until its radius reaches 0 at division. The confinement (c) at which the division occurs, $\frac{L}{2h}\sim10$ for the presented experiment, slightly decreases with the successive divisions. This last observation has yet to be understood. We speculate that the gradient of height of the capillary, which decreases slightly in the area of division, could play a role. More generally for all experiments, the divisions occur for a confinement level in the range $\frac{L}{2h}=[8-20]$, where the disparity amongst experiments can here also be attributed to variations in the imposed height gradients, from one capillary to another.
 
In this series of experiments, the time and spatial resolutions are not large enough to give access to the precise dynamics of a division. Another series of experiments have thus been conducted, using a microscope and a high-speed camera, and focusing on the first division of a long droplet. Because the field of view is limited, to a square of size of typical size $2h$, the position of the neck $l_n$ cannot be quantitatively resolved, but the radius of the neck $r_n$ is measured precisely. The dynamical evolution of $r_n$ for two experiments are presented in Figure~\ref{fig:Dyn}, one conducted at an acquisition frequency of 1000 Hz, with a spatial resolution of 1.692 \mum/pix (orange top oriented triangles), and another conducted at an acquisition frequency of 10000 Hz, with a spatial resolution of 0.840 \mum/pix (red down oriented triangles). 
Figure~\ref{fig:Dyn} also shows the dynamical evolution of $r_n$ for the first division of the droplet in the previous experiment conducted at a frequency of 1 Hz (pink empty circles), and the dynamical evolution of $r_n$ during an experiment without division (dark brown full circles). In the figure, the time $t=0$ corresponds to the division time (evaluated with the temporal resolution of each experiment). For the experiment without division, it is set such that the neck radius match in the converging part of the capillary. Figure~\ref{fig:Dyn} (a) presents the data in a linear plot, while the Figure~\ref{fig:Dyn} (b) presents the same data in a log-log plot. 

The radius of the neck follows three successive regimes; the first one ($t \lesssim -10$s) corresponds to the adaptation of the droplet shape to the confinement gradient, as discussed is the section~\ref{sec:V2}. For droplets that do not undergo division, this regime is reversible when the confinement increases again. In this regime, the evolution of the shape of the droplet is quasi-static, and controlled by the geometry of the problem. The second and third regimes lead to the division of the droplet. Once the droplet enters these regimes, the division always take place. The second regime is very well characterized by a power-law dependence of the neck radius with time $-t$, $\frac{r_n}{h}\gtrsim 0.1$, $\frac{r_n}{h} \sim |t|^{\beta}$ with $\beta \in [1/8-1/6]$ over almost 3 decades in time. For $\frac{r_n}{h}\lesssim 0.1$, the radius of the neck deviates from the latter power law and deepens faster, entering a third regime that also follows an apparent power-law $\frac{r_n}{h} \sim |t|^{\alpha}$ with $\alpha \in [2/3-1]$.

Such power-law behaviours are naturally found in the ultimate fate of the break-up process for a droplet, because of the absence of characteristic length-scale but also in transient regimes~\cite{eggers2008}. The value of the exponent is dictated by which effects dominate and balance amongst surface tension, viscous and inertial forces. Here, the viscosity of the outer fluids dominates, and one expects the simple self-similar form to be broken by the presence of logarithmic terms. Finally, the presence of active stresses is likely to alter the already numerous possible scaling regimes. Investigating such a fascinating question is beyond the scope of the present paper. It would require even much faster acquisition rate, and dedicated experiments. The exponents provided here should be seen as indicative and a source of motivation for future work.

Let us recap our main findings, which we now aim at capturing theoretically, on the basis of the reformulation of the classical Bretherton problem in the realm of active droplets. Two unexpected phenomena have been observed: the convergence of the droplet velocity towards a constant value when the droplet becomes longer than the capillary height, and the spontaneous division of the droplets under high confinement. In the following section, we introduce a simple theoretical framework that will allow to grasp the physics at play behind such observations.

\section{Theoretical approach}

\subsection{Introduction}

The motion of confined droplets or bubbles under the action of an external flow has been widely studied since its original description by F.P. Bretherton in a cylindrical tube~\cite{Bretherton1961} and are usually called Bretherton models in tribute to the British professor. With the emergence of droplet-based microfluidics, a particular interest has been devoted to squared channels~\cite{Baroud2010, Cantat2013}. Pressure, or gravity driven flows are not the only way to induce droplet motions in a channel. Marangoni stresses can also induce the migration of such confined droplets as described and observed in the presence of external thermal gradients~\cite{Mazouchi2000a}. In our case, the motion of the droplets is not externally driven, neither by pressure nor a temperature gradient as there is no observable flow far from the droplet ($U_\infty = 0$), but is the result of local flows induced by the spontaneous establishment of solutes concentration gradients around the droplet. To the best of our knowledge, such a problem has never been considered theoretically before, and is the primary subject of the following section.

The flow around the droplet is driven by a combination of phoretic and Marangoni effects: concentration gradients of all present solute along the interface generates shear stress (denoted $\sigma$) and velocity (denoted $u$) jumps at the interface~\cite{Anderson1989}. For simplicity purpose, we consider in the following that the velocity and stress jumps through the interface result only from the concentration gradient in swollen micelles in the outer fluid, but one should keep in mind that more complex and realistic models of the physico-chemical interactions at the interface have been proposed~\cite{Morozov2020}, taking into account the surfactant concentration at the interface. Then the velocity and stress jumps can be expressed as: $\mathbf{\sigma}_{\textrm{oil},\parallel} - \mathbf{\sigma}_{\textrm{water},\parallel} = -K {\nabla}_\parallel c$ and $u_{\textrm{oil},\parallel} - u_{\textrm{water},\parallel}= M \mathbf{\nabla}_\parallel c$, where $c$ is the solute concentration field, $\nabla_\parallel$ is the gradient operator tangent to the interface, $K \approx k_\textrm{B} T \lambda$ and $M \approx k_\textrm{B} T \lambda^2 / \eta_\textrm{oil}$ where $k_\textrm{B}T$ is the thermal energy, $\lambda$ the typical interaction distance between the solute and the interface and $\eta_\textrm{oil}$ is the oil viscosity. Under the above assumptions, the Marangoni effects dominate with respect to the phoretic ones~\cite{Izri2014}, so that, in what follows, we will assume continuity of the velocity across the interface and the presence of a stress jump at the interface. We further notice that the magnitude of the viscous shear stress in the water phase $\sim \eta_\textrm{water} v / h$ is much smaller than the one in the oil phase $\sim \eta_\textrm{oil} v / e$, where $v$ is a typical velocity in the film and $e$ is a typical lubrication film thickness as the viscosity ratio $\eta_\textrm{water}/\eta_\textrm{oil}\approx 1/40$ and the film thickness to capillary height ratio is small with respect to unity, $e\approx 1$ \mum and $h\approx 100$\mum. Therefore, the tangential stress balance simplifies to $\mathbf{\sigma}_{\textrm{oil},\parallel}= \sigma(x)= -K {\nabla}_\parallel c $. We can thus focus on the flow in the oil phase only, and we refer to the oil viscosity as $\eta$ to lighten the notations. The peculiarity, and difficulty of this problem lies in that the hydrodynamic and the transport of the chemical species (surfactant molecules and swollen micelles) in the solution are non-linearly coupled via the Marangoni stress $\sigma(x)$, which varies along the interface.

\begin{figure}[h!]
\centering
\includegraphics[width=\columnwidth]{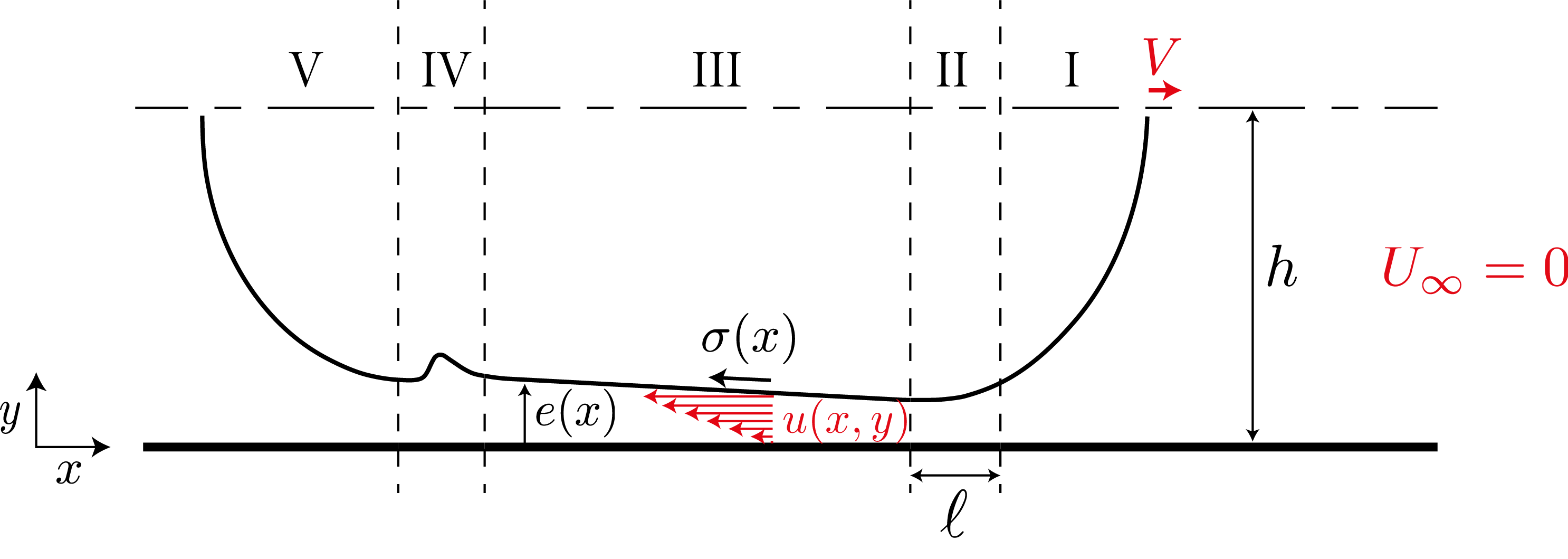}
 \caption{\textbf{Sketch of an elongated droplet in a circular capillary, in the lab frame :} the droplet is divided into five regions, two spherical caps (I) and (V) of radius $h$, two dynamical meniscus, the front dynamical meniscus (II) of typical size $\ell$ and the rear dynamical meniscus (IV) where a neck forms, and a lubrication film (III) of variable thickness $e(x)$, where a Marangoni stress $\sigma(x)$ at the interface induces local flows of velocity $v(x,y)$. The resulting droplet velocity is denoted $V$. Far from the droplet, there is no flow $U_\infty=0$.}
\label{fig:schema_theo}
\end{figure}

The present theoretical description deals with highly elongated droplets, swimming in cylindrical tubes that are axially invariant, leaving aside the case of squared channel~\cite{wong1995a,Mazouchi2000a}. In such confined environment, various lengths of different magnitude are at play : the radius of the capillary $h$, the length of the droplet $L$, the thickness of the lubrication film $e(x)$ and, at the microscopic level, the typical distance of interaction between the solute and the interface $\lambda$~\cite{Anderson1989}. Given the large scale separation between these lengths, $L> h\gg e \gg\lambda$, it is a standard approach to separate the problem in different regions and match the corresponding solutions asymptotically~\cite{Bretherton1961,Mazouchi2000a}. Usually, five zones are distinguished, as exhibited in Fig.~\ref{fig:schema_theo}: the front (I) and rear (V) caps that are supposed spherical, the front (II) and rear (IV) dynamical menisci of variable curvature, and in between the lubrication film (III) that is defined as the limiting solution of the dynamical menisci with a uniform thickness. 

We propose to use here a similar approach. In section~\ref{sec:lubrication}, we derive a lubrication model for the velocity field, coupled to the transport of solute. Then, in section~\ref{sec:lubricatonfilm}, we focus on the lubrication film dynamics (zone III) where capillary flows are negligible behind Marangoni flows and we propose a numerical resolution of the resulting system of equations. This allows to identify the typical scales of the Marangoni stress and the film thickness in the problem. Lastly, in the section~\ref{sec:dynamical_meniscus}, we simplify the equation in the dynamical meniscus by assuming a uniform Marangoni stress, of typical magnitude equal to the one identified in the zone III, and we find a Landau-Levich type equation. The matching of this solution to the spherical cap allows to obtain a scaling relation of the droplet velocity, which we finally compare to the experimental data.

\subsection{Lubrication model in the zone II-III-IV\label{sec:lubrication}}

We consider the steady motion of a water droplet in oil that is assumed to behave as a Newtonian fluid. The Reynolds number $\text{Re} = \rho V h / \eta$, where $\rho$ is the oil density is much smaller than unity so that fluid inertia is neglected. The water-oil interface position, denoted $e(x)$, depends on $x$ and is almost parallel to the direction of motion $x$ in the regions II-III-IV so that one can use the lubrication approximation to describe the flow. Therefore the pressure field $p$ is independent of the normal direction $y$ and is given by the Laplace pressure, via the normal stress continuity: 
\begin{equation}
\label{eq:Laplace}
p(x) = -\gamma e''(x) - \gamma/h,
\end{equation}
where $\gamma$ denotes the water-oil interfacial tension. We have neglected the non-linear terms in the curvature in accordance with the lubrication approximation. We also suppose that the interfacial tension is uniform, on the basis that the surface tension difference resulting from the chemical activities is small (of the order of one tenth) with respect to the equilibrium surface tension ($\gamma_{water/oil-micelles}\simeq 2$ mN/m and $\Delta\gamma_{activity} \sim 0.2$ mN/m). The momentum balance in the direction of motion reduces to $\partial_x p = \eta \partial^2_y u_x$, where $u_x$ is the velocity component in $x$. The shear-stress continuity at water-oil interface gives $\eta \left.\partial_y u_x\right|_{y = e(x)} = \sigma(x)$ at leading order in the lubrication scaling, where $\sigma(x)$ is the Marangoni stress. In the reference frame of the moving droplet, the no-slip boundary condition at the wall reads $\left.u_x\right|_{y = 0} = -V$. As a consequence of the global flux conservation, the flux of water must balance the oil flux, which means that the typical velocity in the film $v$ is of the order of $V h / e $, where $e$ is the typical lubrication film thickness. Hence, the typical velocity in the film is hundred times larger than the drop velocity and we can safely approximate the no-slip boundary condition to $\left.u_x\right|_{y = 0} = 0$. The resulting flow is a linear combination of a Poiseuille and Couette terms and reads:
\begin{equation}
\label{eq:velocity_field_meniscus}
u_x(x,y) = \frac{p'(x)}{2\eta} \bigg(y^2 - 2 y e(x) \bigg) + \frac{\sigma(x)y}{\eta},
\end{equation}
\begin{equation}
\label{eq:velocity_field_y_meniscus}
u_y(x,y) = -\frac{p''(x)}{2\eta}\bigg(\frac{y^3}{3} - y^2e(x) \bigg) + \frac{p'(x)e'(x)y^2 }{2\eta} - \frac{\sigma'(x)y^2}{2\eta}.
\end{equation}

Computing the pressure gradient from~(\ref{eq:Laplace}) and integrating the flow in $y$ allows us to express the flux conservation. We find:
\begin{equation}
\label{eq:dim_bretherton}
\frac{\gamma e^3(x) e'''(x)}{3\eta} + \frac{\sigma(x) e^2(x)}{2\eta} = -\phi,
\end{equation}
where the oil flux per unit of orthoradial length is denoted $-\phi$, such that $\phi$ is a positive quantity. The first term on the left hand side of this equation corresponds to the driving by capillarity and the second one to the Marangoni ones. The global mass conservation in a plane perpendicular to the capillary axis implies that the flux of advected oil in the lubrication layer must balance the longitudinal water flux
\begin{equation}
\label{eq:mass_conservation}
\pi h^2 V =  2\pi h \phi.
\end{equation}
The Marangoni stress originates microscopically from the gradients of swollen micelles interacting with the interface. The transport of swollen micelles obeys the stationary advection-diffusion equation 
\begin{equation}
\label{eq:adv-diff}
u_x(x,y) \frac{\partial c}{\partial x} + u_y(x,y) \frac{\partial c}{\partial y} = D \bigg(\frac{\partial^2 c}{\partial x^2}+ \frac{\partial^2 c}{\partial y^2}\bigg) \approx D \frac{\partial^2 c}{\partial y^2}, 
\end{equation}
where $D$ is the diffusion constant of swollen micelles in solution and $c(x,y)$ denotes the concentration fields of solute. In what follows, the diffusion terms in $x$ are neglected with respect to the ones in $y$ in agreement with the lubrication approximation. The swollen micelles are produced at the water-oil interface with a rate $A$, also called the droplet activity, which is assumed to be constant, and gives the boundary conditions at the water-oil interface
\begin{equation}
\label{eq:flux-nosaturation}
\left. D\frac{\partial c}{\partial y}\right|_{y = e(x)} = A.
\end{equation}
The wall is assumed to be impermeable such that the diffusive flux vanishes at the wall, \textit{i.e.}$\left. -\frac{\partial c}{\partial y}\right|_{y = 0} = 0 $. The Marangoni stress is induced by the concentration gradient tangent to the interface as
\begin{equation}
\label{eq:Marangoni}
\sigma(x) = -K \left. (\vec{t}.\vec{\nabla})c\right|_{y = e(x)} = -K \left. (\partial_x + e'(x)\partial_y)c\right|_{y = e(x)}.
\end{equation}

In the following we shall not solve the general lubrication problem but focus on the dominant swimming mechanism with the aim at identifying the scaling governing the droplet velocity. In the next section, we focus on the solution in the lubrication film, which corresponds to the zone III in Fig.~\ref{fig:schema_theo}. 

\subsection{Scaling and numerical solution in zone III. \label{sec:lubricatonfilm}}
In this region, the water-oil interface is nearly flat, although we stress that the film thickness is not necessarily uniform in the lubrication film zone, which is the major difference with classical Bretherton models. The goal of this section is to provide a solution of the lubrication model in the zone III. 

A first step is to identify the proper length and stress scales at play. The dimensionless ratio $\phi/D$ is the ratio between horizontal and vertical transport scales. According to Eq.~\eqref{eq:adv-diff}, $u_x/D \sim x^*/{e^*}^2$, where $e^*$ and $x^*$ denote the characteristic thickness of the film and $x$ scales, so that $\phi/D \sim x^*/e^*$. This ratio compares advection to diffusion and is therefore analogous to a local P\'{e}clet number. In the experiments presented here, its typical magnitude is large $\text{Pe} = \phi/D \sim 100$ (see Appendix~\ref{sec:AN}).
One finds from equation~(\ref{eq:flux-nosaturation}) a concentration scale $c^*=\frac{A}{D} e^*$ and, from equation~(\ref{eq:Marangoni}), a stress scale $\sigma^*= \frac{KA}{Pe D} = \frac{KA}{\phi}$. The lengthscale $e^*$ is chosen such that the Marangoni driving dominates in equation~(\ref{eq:dim_bretherton}) so that $\frac{\sigma^* {e^*}^2}{\eta} = \phi$, and one obtains:
\begin{equation}
e^* = \phi \sqrt{\frac{\eta}{KA}}, \quad x^* = e^* \phi/D, \quad \sigma^* = \frac{KA}{\phi}, \quad c^* = \frac{\phi}{D}\, \sqrt{\frac{A\eta}{K}},
\end{equation}
Coming back to equation~(\ref{eq:dim_bretherton}), we then find that the Marangoni driving dominates as soon as $\sigma^*{e^*}^2 \gg \frac{\gamma e^{*4}}{x^{*3}}$, or, in other words when $x^* \gg \ell^*$, with $\ell^* = e^* \bigg(\frac{\sqrt{\eta K A}}{\gamma} \bigg)^{-1/3}$. One notices that the length scale $\ell^*$ has a similar scaling form as the dynamical meniscus length in the classical Landau-Levich-Derjaguin-Bretherton problem. 
A nice way to see the analogy is to understand the ratio $\text{Ca} = \frac{\sqrt{\eta K A}}{\gamma}$ as the ratio of two velocities, the Marangoni driving velocity $v^* = \sqrt{\frac{K A}{\gamma}}$ and the capillary one $V_{\gamma}=\frac{\gamma}{\eta}$ exactly as in the classical problem where the driving velocity is externally set and $\text{Ca}_\text{B} = V/V_{\gamma} = \eta V/\gamma$.
In the experimental system, the capillary number is estimated to $\text{Ca} \sim 10^{-3}$ (see Appendix~\ref{sec:AN}), which is small with respect to unity. Therefore it justifies the use of the Bretherton type scale separations in the present work. 

We introduce the dimensionless variables with $\tilde{\cdot}$ as $e(x) = e^* \tilde{e}(\tilde{x})$, $y = e^* \tilde{y}$, $x = x^* \tilde{x}$, $c(x,y) = c^* \tilde{c}(\tilde{x}, \tilde{y})$, $\sigma(x) = \sigma^* \tilde{\sigma}(\tilde{x})$ and the equations~\eqref{eq:dim_bretherton}\eqref{eq:adv-diff}\eqref{eq:flux-nosaturation}\eqref{eq:Marangoni} become:
\begin{equation}
\label{eq:DLadv-diff}
-\frac{2\tilde{y}}{\tilde{e}^2} \frac{\partial \tilde{c}}{\partial \tilde{x}} + \frac{2\tilde{e}'y^2}{\tilde{e}^3}\frac{\partial \tilde{c}}{\partial \tilde{y}} =  \frac{\partial^2 \tilde{c}}{\partial \tilde{y}^2},
\end{equation}
\begin{equation}
\label{eq:DLflux}
\left. \frac{\partial \tilde{c}}{\partial \tilde{y}}\right|_{\tilde{y} = \tilde{e}} = 1, \quad \left. \frac{\partial \tilde{c}}{\partial \tilde{y}}\right|_{\tilde{y} = 0} = 0, 
\end{equation}
\begin{equation}
\label{eq:DLMarangoni}
\tilde{\sigma}(\tilde{x}) = -\frac{2}{\tilde{e}^2(x)} = \left. (\partial_{\tilde{x}} + \tilde{e}'(\tilde{x})\partial_{\tilde{y}})\tilde{c} \right|_{\tilde{y} = \tilde{e}}.
\end{equation}
The later equations are solved numerically using a volume-of-fluid method~\cite{tryggvason2001}. The Eq.~\eqref{eq:DLadv-diff} is analogous to a 1D heat equation where $-\tilde{x}$ represents time. Therefore, we set ‘‘initial conditions'' at $\tilde{x} = 0$ and solve for negative $\tilde{x}$. In this model, the initial conditions represents an arbitrary $x$ position near the boundary between the zone II and III in Fig.~\ref{fig:schema_theo}. There we assume that the solute has not diffused over the full lubrication film and is localized near the water/oil interface. We proceed as follows: we first choose an initial thickness $\tilde{e}(\tilde{x}=0)$ and we take a initial concentration fields as $\tilde{c}(\tilde{x}=0,\tilde{y}) \propto \exp(\frac{\tilde{e}(\tilde{x}=0) - \tilde{y}}{\mathcal{L}})$, where $\mathcal{L}$ is a dimensionless length scale that would correspond to the length over which the solute has diffused on the region I-II in Fig.~\ref{fig:schema_theo}. The prefactor of the initial concentration is set to be consistent with the flux boundary condition Eq.~\eqref{eq:DLflux}.
\begin{figure}[t!]
\centering
\includegraphics[width=0.9\columnwidth]{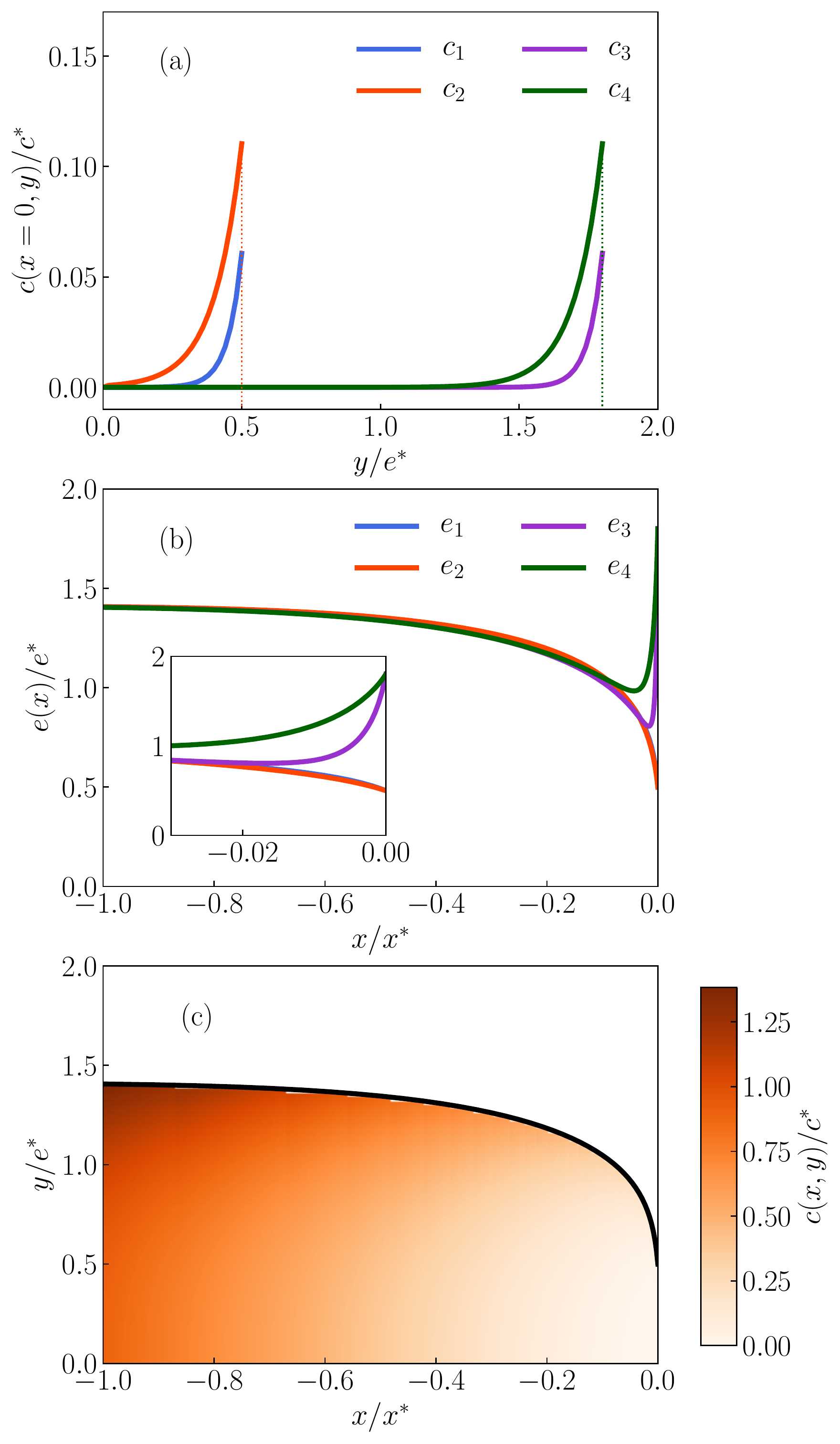}
 \caption{(a) Initial conditions at $x/x^*=0$ for the concentration fields given by $\tilde{c}(\tilde{x}=0,\tilde{y}) \propto \exp(\frac{\tilde{e}(\tilde{x}=0) - \tilde{y}}{\mathcal{L}})$. The initial thickness is set to $\tilde{e}(\tilde{x}=0) = 0.5$ (resp. $1.8$) in the initial conditions denoted $1 - 2$ (resp. $3 - 4$) and the dimensionless length $\mathcal{L} = 0.05$ (resp. $0.1$) in $1-3$ (resp. $2-4$). (b) Evolution of the non-dimensional film thicknesses along the x-axis in the numerical simulation for the different initial conditions. The inset shows a zoom near $x = 0$. (c) Colormap of the non-dimensional concentration field of solute $c(x,y)/c^*$ resulting from the numerical integration with the initial condition $1$. The thickness profile is highlighted in black. }
\label{fig:advection-diffusion}
\end{figure}

Fig.~\ref{fig:advection-diffusion} displays the numerical solution of Eq.~\eqref{eq:DLadv-diff}. We exhibit four solutions with a subscript $i = (1, 2, 3, 4)$ that differs via their initial conditions, plotted in Fig.~\ref{fig:advection-diffusion}(a). Panel (b) shows the evolution of the lubrication film thickness along the $x$-axis for these four different initial conditions. The lubrication film converges toward a uniform solution, with a constant thickness $e_\infty \simeq \sqrt{2}e^* =  \sqrt{2} \phi \sqrt{\eta/(KA)}$, a constant concentration gradient $\partial_x c$ and a Marangoni stress $\sigma_\infty = -\sigma^* = -KA/\phi$, whatever the initial film thickness and concentration. The exact prefactor $\sqrt{2}$ is obtained analytically in Appendix~\ref{appendix:stationnary_solution}, injecting an uniform solution ansatz in Eqs.~ \eqref{eq:adv-diff}, \eqref{eq:flux-nosaturation} and \eqref{eq:Marangoni}. The concentration field of the solution $1$ is displayed in Fig.\ref{fig:advection-diffusion}(c). As already stressed, the film thickness is not uniform; the uniform solution is obtained only for $|x| \gtrsim x^*$, that is when the solute diffusing front reaches the wall at $y = 0$. The concentration fields in this regime is in very good agreement with the uniform solution computed in Appendix~\ref{appendix:stationnary_solution}. At this stage the problem is not closed as the flux $\Phi$ is still unknown.

\subsection{Scaling of the droplet velocity.\label{sec:dynamical_meniscus}}

In this section, we aim at deriving a scaling law for the droplet velocity using the aforementioned scales. Solving the lubrication problem in zone II, requires the full resolution of equation~(\ref{eq:dim_bretherton}), which has no simple or scaling solution. We therefore assume that the Marangoni stress in zone II does not change much and we give it a uniform value set by that of the uniform solution of the lubrication film $\sigma(x) = \sigma_\infty = -\sigma^*$. This is a strong assumption which will only be validated by comparison with the experimental data. Having done so, Eq.~\eqref{eq:dim_bretherton} is written in a closed form and one can then use $e_\infty = \sqrt{2}e^*$ and $\ell^*$ as a thickness and $x$ scale and write the resulting flux conservation in a universal form using the dimensionless variables $E(X) = e(x)/(\sqrt{2}e^*)$, $X = x/\ell^*$ : 
\begin{equation}
\label{eq:dimless_bretherton}
\frac{2}{3}\frac{\textrm{d}^3E}{\textrm{d}X^3} = \frac{E^2-1}{2 E^3}.
\end{equation}
The latter equation admits a trivial solution $E = 1$, which corresponds to the uniform film usually found in Bretherton models and which is identical to the uniform solution found at $|x|>x^*$ in the previous section. We solve Eq.~\eqref{eq:dimless_bretherton} numerically following the standard Landau-Levich approach. We assume an uniform film at $X \to -\infty$, and linearize Eq.~\eqref{eq:dimless_bretherton} as $E = 1 + \epsilon$, where $\epsilon \ll 1$, which gives $\frac{2}{3}\epsilon'''(X) = \epsilon(X)$. The solution compatible with the flat film at $-\infty$ is $\epsilon(X) = \epsilon_0 \exp\bigg((\frac{3}{2})^{1/3} X\bigg)$, where $\epsilon_0$ is an arbitrary constant. We then solve the initial value problem defined by Eq.~\eqref{eq:dimless_bretherton} using a Runge-Kutta scheme of order 4 and with the linear solution as an initial condition. The numerical solution is found to diverge at $X \to \infty$ with a finite second derivative, leading to $\lim_{x \to \infty} e''(x) = \frac{2.125}{e^*} (\frac{\sigma^* e^*}{\gamma})^{2/3}$ for the dimensional variables. The limit curvature must be matched to the curvature of the spherical caps $\frac{1}{h}$ in order to preserve the continuity of the pressure in the region I and II, which yields to the following relationship 
\begin{equation}
\label{eq:film_thickness}
\frac{e^*}{h} = 2.125 \, \bigg(\frac{\sigma^* e^*}{\gamma}\bigg)^{2/3}.
\end{equation}
Note that, although the right hand side term of eq.~(\ref{eq:dimless_bretherton}) differs from the standard Landau-Levich-Derjaguin and Bretherton one, the film thickness scaling law remains of the same form, as $\sigma^* e^*/\gamma = \sqrt{\eta K A}/\gamma = \text{Ca}$. The reason is that the exponent $2/3$ results from the presence of the third order derivative in the left hand side of eq.~(\ref{eq:dimless_bretherton}) and the fact that the asymptotic matching with the spherical caps involves the curvature, hence the second derivative of $e(x)$, two aspects which are common to our problem and the classical one. The product of Marangoni stress and film thickness appears as the relevant traction force that deforms the interface, analogous to $\eta V$ in the standard Bretherton framework. 
Finally, recalling the global mass conservation Eq.~\eqref{eq:mass_conservation}, one finds the swimming velocity $V = 2\phi/h = \frac{\sigma^*e^*}{\eta} \frac{e^*}{h} $. Combining this expression with Eq.~\eqref{eq:film_thickness}, we find:
\begin{equation}
\label{eq:drop_velocity}
V = 2.125\,  \frac{\sigma^* e^*}{\eta} \, \bigg(\frac{\sigma^* e^*}{\gamma}\bigg)^{2/3} \sim \sqrt{\frac{ K A}{\eta}}\bigg(\frac{\sqrt{\eta K A}}{\gamma}\bigg)^{2/3}
\end{equation}
or, in a more compact form,
\begin{equation}
    \frac{V}{v^*} \sim \text{Ca}^{2/3}.
\end{equation}

A first validation of the present scaling relation is that it predicts a swimming velocity which does not depend on the capillary height $h$, as observed experimentally. 
Second, we can compute an estimation of the ratio between the film thickness and capillary radius $e^*/h \sim (\sigma^*e^*/\gamma)^{2/3} \sim 1/100$, that implies a film thickness of the order of $1$ micron which is consistent with experimental observation.
Finally, a numerical evaluation (using the numerical values given in Appendix~\ref{sec:AN}) leads to a droplet velocity in the micron per second range, which is consistent with what is observed in the experiments. 

\subsection{Saturation of the solute}
The above description finds that the interface profile saturates once the solute has diffused over the film, at a position $-x \sim x^*$. We evaluate $\frac{x^*}{2h}\sim0.1$. 
This contrasts with the experimental observation of an increasingly deep neck with increasing confinement, that leads to division for $\frac{L}{2h} \gtrsim 10$. Beside, we observe experimentally that the daughter droplets that have detached themselves at the rear of the main droplet don't swim. This suggests that there is no more fuel for the propulsion - all micelles in solution have been saturated with water. We speculate that the spontaneous division of the droplet is related to this saturation of swollen micelles at the rear of the droplet, an ingredient absent so far from our theoretical description. 

As a matter of fact, one expects the presence of swollen micelles near the water-oil interface to disturb the sorption kinetic of the surfactant molecules and to slow down the emission of swollen micelles~\cite{Morozov2020}. In the model, the lubrication film thickness at large $-x$ is found to scales as $e_\infty \propto 1/\sqrt{A}$, and thus is expected to increase as the emission rate decreases.  A precise description of a physico-chemistry that trigger the saturation is beyond the scope of the paper, but as a minimal description, the model is consistent with the scenario of a growing lubrication film, at the rear of the droplet, where the non-uniformity is now driven by the saturation of swollen micelles. 

Let us simply point out a few elements of thoughts. For droplets that are not too long, we expect a continuous matching between a modest increase of the lubrication film thickness and the rear meniscus, where the active stresses have vanished. In such a case, the dynamics remains steady and the evolution of the droplet shape should be reversible when entering and escaping a constriction zone, as observed experimentally. On the contrary for very long droplets, the diverging lubrication film thickness generates strong curvatures, which will eventually trigger a Rayleigh-Plateau instability and lead to an irreversible dynamical regime the ultimate fate of which is the division of the droplet.

\section{Conclusions}

In this work, we presented first-of-a-kind experimental measurements of the behavior of a swimming droplet in one-dimensional capillaries of different geometries, namely square capillary, circular capillaries and stretched circular capillaries. For high enough confinement, the velocity of the droplet converges toward a small but non-zero value, while the lubrication layer, which separates the droplet from the wall, becomes of non-constant thickness and a neck forms at the rear of the droplet. Under continuously increasing confinement, the deepening of the neck is observed to lead to successive spontaneous divisions of the droplet. A brief study of its dynamic shows a rich behaviors that can be the ground for future works. 

We introduce a simplified model for the motion of such a confined droplet following the standard Bretherton approach, with the major difference that the flow is locally driven by solute concentration gradient at the interface of the droplet. We focus on the front dynamical meniscus and the lubrication layer. The latter is treated using the lubrication layer approximation, and we find that the solute concentration converges toward a uniform solution far from the front meniscus with a uniform thickness. The front dynamical meniscus is only treated partially, simplifying the transport equation and assuming a uniform stress at the droplet interface. The matching of these two regions, using the aforementioned uniform solution, allows us to find a scaling relation for the emerging velocity of the droplet, which, as observed experimentally, does not depend on the confinement. Finally we argue that the saturation of the swollen micelles at the rear of the droplet, decreases the solute emission flux, giving rise to increasing film thickness, which ultimately is prone to induce the spontaneous division of long enough droplets.

As the theoretical approach presented in this work was meant to be kept simple, a certain number of hypothesis have been used. Among them, the assumption of a uniform Marangoni stress in the dynamical meniscus is the strongest one. Ideally one would need to solve the advection-diffusion problem also in this region to find the precise prefactor for the droplet velocity and check the robustness of the scaling law derived here. 

The experimental measurement of the flow field around a confined swimming droplet, Figure~\ref{fig:PIV} (b) shows that the hydrodynamics in front of the droplet is also more complex that what we considered theoretically. More specifically, we observe a large re-circulation area, which breaks the axisymmetry of the problem. How to capture this symmetry breaking and coupling it to the above description is a completely open question. Not only does it most likely alter the droplet velocity but is also bounded to have consequences on the interactions between two droplets. 

Finally, the spontaneous division of the droplet under increasing confinement is an unexpected consequences of the limited amount of empty reverse micelles in solution. In this work, we kept the initial concentration of micelles constant. In a complex environment where the concentration of reverse micelles could vary with time and space, this instability would be triggered only in region where the "food" is scarce, an amazing behavior to observe, especially in the perspective of using simple physical systems in the design of probiotic systems. 

\section*{Conflicts of interest}  
There are no conflicts to declare.

\section*{Contributions}
Charlotte de Blois and Olivier Dauchot initiated the experimental project.
Charlotte de Blois and Saori Suda carried out the experiments overseen by Mathilde Reyssat. Vincent Bertin initiated the theoretical part and performed the analytical calculations and the numerical simulations. All authors participated in the development of the theory, discussed the results and contributed to the final manuscript. Charlotte de Blois and Vincent Bertin wrote the manuscript under the guidance of Mathilde Reyssat and Olivier Dauchot.

\section*{Acknowledgements}
We thank Sébastien Michelin for interesting discussions all along the duration of the present projects and other related ones. Charlotte de Blois was sponsored by a doctoral fellowship from the Ecole Doctorale Physique en Ile de France. Saori Suda was sponsored by The Kyoto University Foundation (Public Interest Incorporated Foundation) and by JSPS KAKENHI Grant Number 20J15804. 

\appendix  

\section{Material and Methods\label{sec:MM}}

The experimental system is a water droplet inside a glass capillary filled with a continuous oil-surfactant phase consisting of a surfactant mixed in squalane. The surfactant is the mono-olein, a nonionic surfactant at a concentration $c=$ 25 mmol/L, which is far above its critical micellar concentration (CMC $\simeq$ 5 mmol/L). The droplets are produced using a \copyright Femtojet apparatus by injecting a single droplet of controlled size in the micro-channel previously filled with the oil-surfactant solution, and left opened at both ends. The length of the droplet formed varies between 0.25 and 8 times the capillary inner size (for reference, it would correspond to equivalent spherical droplets of radius between 25 \mum and 250 \mum.)
The droplets are made from a (milli-Q) water solution of $15$\%wt NaCl. The continuous phase is a $25$ mM mono-oleine surfactant (MO; 1-oleoylrac-glycerol, $99$\%, Sigma-Aldrich) solution in squalane (Sq; 99\%, Sigma-Aldrich). The room temperature is kept above $25^{o}C$ in order to avoid mono-oleine crystallization~\cite{Qiu2000}.
 
Three different 1D geometries are used:
\begin{enumerate}
\item Square glass capillaries (Figure~\ref{fig:Shape} (a)) of length 5 cm, and of four different inner sizes: $2h=400$ \mum, $2h=200$ \mum, $2h=100$ \mum and $2h=80$ \mum. The capillaries are either used native, or silanized beforehand. $h$ is then defined as half the inner dimension of the capillary.
\item Circular glass capillaries (Figure~\ref{fig:Shape} (b)) of length 10 cm, and of two different inner sizes: $2h=200$ \mum and $2h=100$ \mum, all silanized. $h$ is then defined as the radius of the capillary. To make possible the imaging through the curved shape of these capillaries, the observation section is immersed into glycerol whose refractive index is close to glass. 
\item Stretched circular capillaries (Figure~\ref{fig:El}) of length 3-5 cm, whose inner radius varies continuously along their length between $2h=100$ \mum (at both ends), and a constriction of diameter $2 h_\text{min}$, in the middle of the capillary, with a typical gradient of diameter $\frac{d h}{d x}=\pm 0.02$. Thus they present a convergent region followed by a divergent one. These capillaries are designed from circular glass capillaries of inner size $2h=100$ \mum that are stretched by hand by locally heating and stretching a portion of the capillary of typically 0.5 cm. These stretched capillaries are silanized.
\end{enumerate}

Three sets of experiments are conducted. 
\begin{enumerate}[label=\roman*.]
\item For the first set of experiments, images of a droplet inside a square or circular capillary are acquired using a AZ100 Nikon macroscope, equipped with x1 air objective. The camera is a black and white camera Dalsa Falcon II, with a resolution of 4096 x 3072 pixels, and an acquisition frequency of 1 Hz. The macroscope has a continuous zoom between x1 and x8, and thus has a variable resolution, which is measured before each experiment by using a calibration slide. Typically, to visualize an area of 1 cm in diameter, we use the x3 zoom, which gives a resolution of 0.3 pix/\mum. The droplet motion in the capillary is then tracked in the frame of reference of the laboratory, and its shape is detected using an intensity threshold algorithm. 

\item A second set of experiments is conducted to measure the flow field around a droplet inside a circular capillary, using a Particle Image Velocimetry (PIV) technique. Red fluorescent colloids tracers (Fluoro-Max$^{TM}$, 0.6 $\mu$m Red Fluorescent Polymer Microspheres, Thermo scientific) are added in the oil phase. The seeding is set to approximately $0.25$ colloids/\mum$^3$, which corresponds in an illumination plane of depth 5 \mum to little more than one colloid per \mum$^2$, or one colloid per two pixels$^2$. The images are acquired with a CCD camera (Andor Zyla 5.5) in the median plane of the droplet using confocal microscopy with a x10 objective, and a laser beam at 540 nm, which is the absorption wavelength of the tracers. The acquisition frequency is $10$ frames/s and the exposure time is $50$ ms. The spatial resolution in the plane is $0.65\mu$m/pixel. For each experiment, $100$ images of the droplet and the surrounding flow field are acquired. 
The PIV analysis is performed using the PIVlab~\cite{Thielicke2014} code on \copyright Matlab. Pre-processing is done using a Wiener filter of window size 3 pixels. Then the PIV is performed by using cross-correlation between two successive images in two passes of respective interrogation areas of 64 pixels and 32 pixels,  (which corresponds typically to a window containing ten tracer particles) and with a window overleap of 50\%. The walls and the inside of the droplet are excluded from the PIV by designing a moving mask for each image. The droplet's mask is designed to be slightly smaller than the droplet size, and is moving with the droplet. Post-processing validation is not used. The final spatial resolution of the mapping of the flow field is then 16 \mum/pixel.
This provides us with the velocity field in Cartesian coordinates attached to the lab frame at each time step. We then average in time the instantaneous flow fields obtained from PIV, thereby reducing the experimental noise. \\

\item Finally, a third set of experiments focus on the dynamics of the rear of the droplet. Images are acquired using a Leica microscope equipped with a x10 air objective, and a fast camera Photron Fastcam SA3 with varying acquisition frequencies between 1000 Hz and 10000 Hz. The same image processing than for the first set of experiments is used to detect the droplet interface. 
\end{enumerate}

\section{Numerical applications \label{sec:AN}}

In this section, we give the numerical values used to do the numerical applications done in the main text which are based on the ones used in~\cite{Izri2014}. 

\textbf{Peclet number :} using a swollen micelle radius of \mbox{$\delta$=2 10$^{-9}$ m}, the oil viscosity $\eta = $  40 10$^{-3}$ Pa.s, the diffusion constant is evaluated as \mbox{$D = \frac{k T}{6\pi\eta \delta}\sim 10^{-12}\text{m}^2/\text{s}$}, \mbox{$k_B$=1.38 10$^{-23}$ J/K} is the Boltzmann constant and \mbox{$T$= 300 K} is the temperature. We compute $\phi = \frac{Vh}{2} = 1.5~10^{-10}\text{m}^2/\text{s}$ using the experimental parameters in Fig.~\ref{fig:V}, $V=3$ \mum/s being the velocity of the droplet and $h=50$ \mum is the height of the channel. Then $\text{Pe} = \phi/D \sim 100$. 

\textbf{Capillary number :} the surface tension of the water-oil interface is $\gamma = 1.7 \, 10^{-3}$ Pa.m, measured using the pendant drop method~\cite{Berry2015}. The Marangoni constant $K$ is derived from the relation \mbox{$K=k_B T \lambda $}, where \mbox{$\lambda$ =10 10$^{-9}$ m} is the typical distance of interaction between the solute and the interface. Then \mbox{$K \sim 10^{-29}\text{J.m}$}. The activity, or surface flux $A$, is derived from the relation \mbox{$A=\frac{3}{4 \pi}\frac{\kappa}{\delta^3}$}, where \mbox{$\kappa$=5 10$^{-8}$ m/s} is the decrease rate of the radius of an unconfined droplet. Then $A \sim 10^{18} \text{m}^{-2}.\text{s}^{-1}$. Finally, the capillary number can be evaluated as $\text{Ca} = \frac{\sqrt{\eta K A}}{\gamma} \sim 10^{-3}$.

\section{Uniform solution}
\label{appendix:stationnary_solution}
In this section, we write the stationary solution of the solute transport equations. We make the following ansatz for the concentration field and thickness evolution 
\begin{equation}
\tilde{c}(\tilde{x},\tilde{y}) =A_0 + A_1 \tilde{x} + A_3 \tilde{y}^3, \quad \quad \tilde{e}(\tilde{x}) = E. 
\end{equation}
Injecting this solution in Eqs.~ \eqref{eq:adv-diff}, \eqref{eq:flux-nosaturation} and \eqref{eq:Marangoni}, one find the following coefficient
\begin{equation}
A_1 = -1, \quad A_3 = \frac{1}{6}, \quad E = \sqrt{2},
\end{equation}
and $A_0$ is a free parameter.



\balance


\bibliography{SM1D} 
\bibliographystyle{rsc} 

\end{document}